\documentclass[aps,prb,reprint,showpacs,amsmath,amssymb,superscriptaddress,floatfix,longbibliography]{revtex4-1}
 \usepackage{float}

\usepackage{bm}
\usepackage{graphicx}
 \newcommand\diag{\operatorname{diag}}

\begin{document}
\title{Local impurity in multichannel Luttinger liquid}

\author{V. Kagalovsky}
\affiliation{Shamoon College of Engineering, Beer-Sheva 84105, Israel}
\author{I. V. Lerner}
\affiliation {University of Birmingham, School of Physics \& Astronomy, B15 2TT, UK}
\author{I. V. Yurkevich}
\affiliation {Aston University, School of Engineering \& Applied Science, Birmingham B4 7ET, UK}

\begin{abstract}
We investigate the stability of conducting and insulating phases in multichannel Luttinger liquids with respect to embedding a single impurity. We devise a general approach for finding critical exponents of the conductance in the limits of both weak and strong scattering. In contrast to the one-channel Luttinger liquid, the system state in  certain parametric regions depends on the scattering strength which results in the emergence of a bistability.   Focusing on the two-channel liquid, the method developed here enables us to provide a generic analysis of phase boundaries governed by the most relevant (i.e.\ not necessarily single-particle) scattering mechanism. The present approach is applicable to channels of different nature as in fermion-boson mixtures, or to identical ones as on the opposite edges of a topological insulator. We show that interaction per se cannot provide protection in particular case of topological insulators realized in narrow Hall bars.
\end{abstract}

\maketitle

\section{Introduction}
The recent advances in study of topological insulators have led to a wider search for non-Abelian states in condensed matter systems and brought to life a set of effective theories describing such exotic states. One of the promising models capable of catching the essential physics of non-Abelian quantum Hall states is an anisotropic system consisting of array of coupled one-dimensional (1D) wires  \cite{TK2014}. This model was used for construction of integer  \cite{Sondhi2001} and fractional quantum Hall states  \cite{Kane2002}. Sliding phases in classical $XY$ models  \cite{OHern1999}, smectic metals  \cite{Vishwanath2001} and many other exotic states are all described by the sliding Luttinger liquid (LL)  model  \cite{MKL2001}. In general, in these and other models of multichannel LL of  translationally invariant (clean) systems, interactions may only open a gap blocking some degrees of freedom and leading to new gapless states for the remaining gapless excitations.

This is one of the reasons of focusing research interest on a multichannel LL with translational invariance broken by a single or multiple impurities. Many specific studies of various two-channel LL with broken translational invariance, like  1D binary cold-atomic mixtures, \cite{HoCaz,*FB-Mat&&LukDem,BF-Simon:10,*BF-Simon:12} electron-phonon LL, \cite{martin05,GYL:2011,*GYL:2011a,*YGYL:2013,IVY:2013,*YY2014} or topological insulators with impurity scattering between opposite edge currents,  \cite{PRL09-Chamon,Santos2015}  have been based on the seminal renormalization group (RG) analysis  \cite{KaneFis:92a,*KaneFis:92b,*KF:92b} of the impact of a single impurity on the conductance of a  single-channel LL. This analysis shows that  such an impact is fully governed by the value of the Luttinger parameter $K$. At  temperatures $T\to 0$ the LL becomes a complete insulator  for any strength of backscattering from impurity for $K<1$ (fermions with repulsion),    or  behaves as translationally invariant LL (i.e.\ becomes an ideal conductor \cite{MasStone:95,*Ponomarenko:95,*SafiSch:95})  for $K>1$ (bosons with repulsion or fermions with attraction). All these examples were special: for chiral currents on the opposite edges of a topological insulator the Luttinger parameters were the same  \cite{PRL09-Chamon,Santos2015} while there was no intra-channel interaction in one of the channels (fermions in the binary cold-atomic mixtures or phonons in the electron-phonon LL) in other examples of the two-channel LL  \cite{HoCaz,*FB-Mat&&LukDem,BF-Simon:10,*BF-Simon:12,martin05,%
GYL:2011,*GYL:2011a,*YGYL:2013,IVY:2013,*YY2014}.

In this paper we develop a general formalism for the RG analysis of the impact of a single impurity on the conductance of an multichannel LL. The results are also applicable to a disordered multichannel LL at moderate temperatures when the thermal length is smaller than mean distance between impurities  -- the limit opposite to that required for the Anderson or many-body localization  \cite{GS1988,BAA1,*BAA2,*AlAlS:2010}. We show that the RG dimensions are governed by a real symmetric $N\times N$ `Luttinger' matrix ${\mathsf{K}}$ whose diagonal elements are defined via the Luttinger parameters and velocities in each channel while off-diagonal ones involve the inter-channel interaction strengths.

We apply the formalism to analyze in detail the conductance of a two-channel LL with arbitrary parameters as well as easily reproduce known results  \cite{PRL09-Chamon} for  scattering between two opposite edge states in a topological insulator.  The  LL can be built, e.g., from the binary cold-atomic mixtures where the values of the Luttinger parameters  in each channel can be arbitrary. Although the RG flows are  governed by ${\mathsf{K}}$, this might be not sufficient to define the conducting state of the LL: it is possible that for the same values of the elements of ${\mathsf{K}}$ both the conducting and insulating channels are \emph{stable}
 against embedding the impurity, signalling the existence of an unstable fixed point with the RG flows in its vicinity governed by the impurity scattering strength. We also discuss the possibility of the two-  or multi-particle scattering from the impurity becoming more relevant than a single-particle one in a certain parametric interval  \cite{KLY:16d}.   In this case there exists a region of parameters where both insulating and conducting phases become \emph{unstable}, indicating the existence of an attractive fixed point or the possibility to construct different initial channels.

\section{Model}
We consider a generic multichannel Luttinger liquid  with inter-channel interactions but only intra-channel scattering from impurities, focusing on the two standard limits of a weak scattering (WS) from the impurity or a weak link (WL) connecting two clean semi-infinite channels.   The conductance of an ideal single-channel LL is known   \cite{MasStone:95,*Ponomarenko:95,*SafiSch:95} to be equal to $e^2/h$ independently of the interaction strength parameterized with the Luttinger parameter $K$.  The RG analysis of the impurity impact  (whether it is WS or WL)   on the conductance $G$ shows  \cite{KaneFis:92a,*KaneFis:92b,*KF:92b} that it is fully governed by $K$: when the temperature tends to zero $G$ vanishes for $K<1$ or goes over to the ideal limit, $e^2/h$,  for $K>1$.

We will show that a phase diagram for the multichannel LL can be drastically different, with the emergence of a region where for a given set of Luttinger parameters the limiting conductance of some or all channels might depend on the scattering strengths. The pivotal role in determining the conducting properties of the multichannel LL is played by the Luttinger matrix $\mathsf{K}$ that generalizes $K$. Its form does not depend on the scattering so that we start with defining  $\mathsf{K}$ in the clean limit.

\subsection{Multichannel Luttinger liquid}
The low-energy Hamiltonian of the usual single-channel LL can be written in the Haldane representation \cite{HALDANE:81} as \begin{align}\label{H}
{\hat H}_1&=\frac{v}{2\pi}\int dx \left[\frac{1}{K}(\partial_x{\hat\theta})^2+K(\partial_x{\hat\varphi})^2\right]\,,
\end{align}
where $v$ is the velocity, $K$ is the Luttinger parameter, and the canonically conjugate operators $\hat{\theta }$ and $\hat{\varphi }$ describe correspondingly the density fluctuations, $\delta{\hat{n}}=\partial_x{\hat{\theta}}/\pi$, and current, ${\hat j}={\partial_x\hat\varphi}/\pi$,
 and obey the commutation relation
\begin{equation}\label{comms}
\Big[{\hat\theta}(x)\,,{\hat\varphi(x')}\Big]=\frac{i\pi}{2}\, \operatorname{sgn} (x-x')\,.
\end{equation}
 The corresponding Lagrangian density, $\mathcal{L}_1$,  can be written in matrix notations as
\begin{equation}
\mathcal{L}_1=\frac{1}{2\pi}(\theta,\,\varphi)\left[\tau_1 \,\partial_t+
                                                                 \begin{pmatrix}
                                                                   vK^{-1} & 0 \\
                                                                  0 & vK \\
                                                                 \end{pmatrix}
                                                               \partial_x
\right] \partial_x\!
                                                                       \begin{pmatrix}
                                                                         \theta \\
                                                                         \varphi \\
                                                                       \end{pmatrix}
\end{equation}
where $\tau_1$ is the Pauli matrix and $\theta$ and $\varphi $ are the bosonic fields corresponding to the operators $\hat{\theta}$ and $\hat{\varphi  }$.

The Lagrangian density of the $N$-channel LL with channels coupled only by interactions can be represented in a similar way as
\begin{equation}\label{LN}
\mathcal{L}=\frac{1}{2\pi}({\bm \theta}^{\rm T},\,{\bm \varphi}^{\rm T})\left[\tau_1 \,\partial_t+
                                                                  \begin{pmatrix}
                                                                 {\mathsf{V}}_{\theta} & 0 \\
                                                                  0 & {\mathsf V}_{\varphi} \\
                                                                \end{pmatrix}
                                                               \partial_x
\right] \partial_x\!
                                                                       \begin{pmatrix}
                                                                         \bm\theta \\
                                                                         \bm\varphi \\
                                                                       \end{pmatrix}
\end{equation}
where the density fluctuations and currents in each channel are combined to form the vectors
\begin{equation}
{\bm \theta}^{\rm T}=(\theta_1,\,\theta_2, ..., \theta_N)\,;\quad
{\bm \varphi}^{\rm T}=(\varphi_1,\,\varphi_2, ..., \varphi_N)\,.
\end{equation}
The cross-terms  $\propto \partial_x\bm\varphi^{\mathrm T}\cdot\partial_x\bm\theta$, are absent since they would break inversion symmetry. The diagonal elements of the real symmetric density-density  and current-current interaction matrices, ${\mathsf V}_{\theta}$ and ${\mathsf{V}}_{\varphi}$,
\begin{equation}
{  V}^{ii}_{\theta}=\frac{v_i}{K_i}\,,\quad {  V}^{ii}_{\varphi}=v_i\,K_i\,,
\end{equation}
account for intra-channel interactions. They are parameterized  by the (renormalized) velocities, $v_i$,  and the Luttinger parameters, $K_i$,  in each channel. The inter-channel interactions are accounted for by the off-diagonal matrix elements ${ V}_{\theta}^{ij} $ and ${  V}_{\varphi}^{ij} $ of  ${\mathsf{V}}_\theta$ and ${\mathsf{V}}_\varphi $.

{Sometimes it is   convenient to represent the Lagrangian in terms of the fields ${\bm{\varphi }}_{{\mathrm{R,L}}} $ of chiral left- and right-movers, which are related to ${\bm{\theta}}$ and ${\bm{\varphi }}$ by the standard rotation}
\begin{align}\label{chiral}
  {{\bm \theta}}&=\tfrac{1}{2}\big({{\bm \varphi}}_\mathrm{R}-{{\bm \varphi}}_\mathrm{L}\big)\,,&
 {{\bm \varphi}}&=\tfrac{1}{2}\big({{\bm \varphi}}_\mathrm{R}+{{\bm \varphi}}_\mathrm{L}\big)\,.
\end{align}
{Generalizing the usual $g$-ology notations, we denote the $i$-$j$ channel interactions of the density components of the same chirality  as $V_4^{ij}\propto \widetilde{g}_4^{ij}  $, and of the opposite chirality as $V_2^{ij} \propto \widetilde{g}_2^{ij}  $. The rotation, \eqref{chiral}, leads to the  relation   $V^{ij}_{\theta,\varphi }\propto \widetilde g_4^{ij}   \pm \widetilde{g}_2^{ij} $ which will be useful later on.}

To diagonalize the Lagrangian,  \eqref{LN},  we first transform the fields $\bm \theta$ and $\bm \varphi $ as follows
\begin{equation}\label{M}
{\bm {\theta}}={\mathsf M}\,{\widetilde{\bm \theta}}\,,\quad
{\bm \varphi}=({\mathsf M}^{\mathrm T})^{-1}\,{\widetilde{\bm \varphi}}\,,
\end{equation}
so that the commutation relations similar to those in \eqref{comms} between different the components of the corresponding operators are preserved. Then it is convenient \cite{IVY:2013,*YY2014} to choose the matrix ${\mathsf{M}}$ in such a way that the two interaction matrices are reduced to the same diagonal velocity matrix $\mathsf u=\operatorname{diag}({u_1,\ldots,u_N})$:
\begin{align}\label{d2}
  \mathsf{M}^{\rm T}\mathsf{V}_{\theta}\mathsf{M}=
  \mathsf{M}^{-1}\mathsf{V}_{\varphi}(\mathsf{M^{\mathrm T}})^{-1} & =\mathsf{ u}\,.
\end{align}
Introducing the matrix ${\mathsf{K}} \equiv {\mathsf{MM^{\mathrm T}}}$, we rewrite this transformation as follows:
\begin{align}\label{K}
  \mathsf{K}\,\mathsf{V}_{\theta}\,\mathsf{K} &=\mathsf{V}_{\varphi}= {\mathsf{MuM^{\mathrm T}}}\,.
\end{align}
The representation of form  ${\mathsf{B=KAK}}$ exists for any two positive-definite  real symmetric matrices  ${\mathsf{A}}\equiv \{{a^{ij} }\} $ and ${\mathsf{B}}\equiv \{{b^{ij} }\} $.
In particular, for $2\times 2$ matrices  ${\mathsf{K}}$ is expressed via matrix elements of ${\mathsf{A}}$ and ${\mathsf{B}}$  and  {$\kappa \equiv \det {\mathsf{K}}=\sqrt{\det {\mathsf{B}}/\det{\mathsf{A}}}$} as  follows\cite{KLY:16a}
\begin{align}\label{K1}
    {\mathsf{K}}&=\sqrt{\frac{\kappa}{ac-b^2}}\begin{pmatrix}
                                                a & b \\
                                                b & c \\
                                              \end{pmatrix},&& \left\{ \begin{array}{rcl}
                                                                         a & = & b_{11}+\kappa a_{22}   \\
                                                                         b & = &   b_{12}-\kappa a_{12} \\
                                                                         c & = & b_{22}+\kappa a_{11}
                                                                       \end{array}
                                               \right..
\end{align}

The Lagrangian density in terms of the new fields, \eqref{M}, is given by
\begin{align}\label{L tilde}
    \mathcal{L}=\frac{1}{2\pi}(\widetilde{{\bm \theta}}^{\,\rm T},\,\widetilde{{\bm \varphi}}^{\rm T})\Big[\tau_1 \,\partial_t+\tau_0{\mathsf{u}}\,
                                                               \partial_x
\Big] \partial_x\!
                                                                       \begin{pmatrix}
                                                                         \widetilde{\bm\theta} \\
                                                                         \widetilde{\bm\varphi} \\
                                                                       \end{pmatrix},
\end{align}
where $\tau_0$ is the block-diagonal unit matrix in the $\widetilde{\bm \theta}$-$\widetilde{\bm \varphi }$  space and ${\mathsf{u}}$ is the velocity vector, \eqref{d2}. This can be finally diagonalized by rotating to the chiral fields,
introduced similar to \eqref{chiral},
resulting in the Lagrangian density
\begin{align}\label{L-diag}
    \mathcal{L}&= \sum_{\eta=\pm1}  \frac{\eta}{4\pi} \widetilde{\bm \varphi} _\eta^{\mathrm T}\partial_\eta\partial _x\widetilde{\bm \varphi} _\eta\,,&\partial _\eta\equiv \partial _t +\eta{\mathsf{u}}\partial _x,
\end{align}
where $\eta=\pm1$ labels the fields of the right- and left-movers.

{As we consider  a local impurity that leads to intra-channel backscattering within the original channels, we will need the correlation functions of the original fields ${\bm{\varphi }}$ and ${\bm{\theta}}$ to describe its impact.
To find them we start in Section \ref{Sec3} with  the straightforward correlations of  $\widetilde{{\bm{\varphi }}}$ and $\widetilde{{\bm{\theta}}}$ governed by the multichannel LL Lagrangian in diagonal form, \eqref{L-diag},
and use \eqref{M} to transform  back to the original fields.} Then we will show  that it is  the matrix ${\mathsf{K}}$, \eqref{K}, rather than the diagonalizing matrix ${\mathsf{M}}$, that governs the RG flows for the conductance of the multichannel LL in the presence of the local impurity.

 \subsection{Intra-channel scattering}
The RG  analysis \cite{KaneFis:92a} of the impact of a local impurity embedded into a  single-channel LL  was actually the analysis of stability of the initially continuous channel (which has   ideal  conductance $e^2/h$    \cite{MasStone:95,*Ponomarenko:95,*SafiSch:95} for any value of $K$) against embedding a weak scatterer, and of stability  of the initially split (and thus insulating) channel against connecting its two parts by a weak link.

In what follows we represent  initially continuous or split channels of the multichannel LL by  boundary conditions for  $\theta$ and $\varphi $  at the point $x=0$ where a WS or WL will be inserted.
To  treat both the insulating and conducting limits on equal footing we parameterize the boundary conditions in terms of the jumps at $x=0$, $\Delta\theta( t)\equiv \theta({+0,t})-\theta({-0,t})$ and $\Delta\varphi ( t)\equiv \varphi ({+0,t})-\varphi ({-0,t})$,  as follows:
\begin{align}
\label{bc2}
\Delta\theta( t)&=0 \,, &
\Delta\varphi(  t) &=-2\xi\,\theta( 0,t)\,.
\end{align}
Here the limit $\xi\to 0$ represents a  continuous channel and  $\xi\to\infty$   represents a split channel for which there is no current across the split so that $\theta$ vanishes on both its sides while the values  $\varphi({+0}) $   and $\varphi ({-0})$ are mutually independent.

The RG analysis  \cite{KaneFis:92a} shows that the continuous channel is stable against embedding a WS, $\mathcal{L}_{\rm ws }\sim \sum_{  n } v_{n}^{{\mathrm{bs}}} {\mathrm{e}}^{2i{  n}  \theta  }$,  for $K>1$ while the split one is stable against embedding a WL, $\mathcal{L}_{\rm wl}\sim \sum_{  n }v_n^{{\mathrm{tun}}}  {\mathrm{e}}^{i{  n}  \Delta\varphi   }$, for $K<1$.

{The boundary conditions, \eqref{bc2}, are generalized for the multichannel case as}
\begin{align}\label{bc}
\Delta\bm\theta( t)&=0 \,, &
\Delta\bm\varphi( t) &=-2\Xi\,{\bm \theta}(0,t)\,,
\end{align}
where $
\Xi  \equiv\operatorname{diag}(\xi_1, \xi_2, ..., \xi_N)$.  In the final answers  we shall take the physical limit  (denoted below  as $\lim_{\xi }$) in which $\xi_i\to0$ for all the continuous channels   and $\xi _j\to\infty$ for all the split channels.

Our aim is to analyze the RG stability of the boundary conditions in \eqref{bc} with respect to inserting a WS (at $x=0$) into each continuous channel  (where $\xi_i \to0$ ), or inserting a WL into each split channel ($\xi_j \to\infty$). We assume that neither WS nor WL leads to inter-channel scattering. This assumption encompasses most relevant cases of carriers with different spins (e.g., helical channels in topological insulators), or different species (e.g., fermion-boson mixtures), or spatially separated edge currents.
Under this assumption the Lagrangian density of the corresponding local perturbation can be written in a uniform way as
\begin{equation}\label{pert}
  \mathcal{L}_{\rm sc}= \sum_{\bm n }\,v_{\bm n_{\mathrm{bs}},{\bm{n}_{\mathrm{tun}}}}\,{\mathrm{e}}^{2i{\bm n}_{\rm bs} \bm{\cdot\theta }({t})+i{\bm n}_{\rm tun} \bm\cdot\Delta\bm{\varphi }({t})} + {\rm c.c.}
\end{equation}
{{Here $v_{\bm n_{\mathrm{bs}},{\bm{n}_{\mathrm{tun}}}}$ is an amplitude of backscattering in continuous channels or tunneling through split channels with multiplicity of each process characterized by vectors ${\bm n}_{\rm bs}$ and ${\bm n}_{\rm tun}$, respectively, where the former has integer components in continuous and zero in split channels, while the latter integer in split and zero in continuous channels.}}

It is convenient to reformulate the boundary conditions, \eqref{bc},   in terms of the $in-$ and $out-$ chiral fields connected by an ${{S}}$-matrix, ${ {\bm \Psi}}_{\rm out}= {{ {\mathsf{S}}}}\,{ {\bm \Psi}}_{\rm in}$:
\begin{align}\label{in-out}
      { {\bm \Psi}}_{\rm out}({t})&\!=\!
                              \begin{pmatrix}
                                { {\bm \varphi}}_\mathrm{R}(+0,t) \\
                                { {\bm \varphi}}_\mathrm{L}(-0,t) \\
                              \end{pmatrix}\!
                            ,&
   { {\bm \Psi}}_{\rm in}({t})&\!=\!
                              \begin{pmatrix}
                                { {\bm \varphi}}_\mathrm{L}(+0,t) \\
                                { {\bm \varphi}}_\mathrm{R}(-0,t) \\
                              \end{pmatrix}
                             ,
\end{align}
where $\bm \varphi _\mathrm{R,L}\equiv\bm \varphi \pm \bm \theta $, and the ${{S}}$-matrix is given by
\begin{align}\label{S}
{ {{ {\mathsf{S}}}}}&=\begin{pmatrix}
 {{\mathsf{{ R}}}} &  {{\mathsf{{ T}}}} \\
 {{\mathsf{{ T}}}} &  {{\mathsf{{ R}}}} \\
\end{pmatrix},&
   {{\mathsf{{ T}}}}={\mathsf{1}}- {{\mathsf{{ R}}}}&= \left({{\mathsf{1}}+{ \Xi}}\right)^{-1} \,,
\end{align}
with ${\mathsf{R}}$ and ${\mathsf{T}}$ being diagonal matrices made of reflection and transmission coefficients in each channel. In the physical limit,
 \begin{align}\label{P}
   \mathrm{lim}_\xi  {\mathsf{R}}&=\mathsf{P}_\mathrm{c},&  \mathrm{lim}_\xi {\mathsf{T}}&=\mathsf{P}_\mathrm{i},
 \end{align}
where ${\mathsf{P}}_\mathrm{c(i)}$ is the projector  onto the subspaces of  continuous (split) channels, i.e.\ the diagonal matrix whose elements equal $1$ for the conducting and $0$ for the insulating channels (or vice versa).

The scattering and tunneling multiplicity vectors in \eqref{pert} can be formally represented via these projectors as $
{\bm n}_{\rm bs}={\mathsf{P}}_{\mathrm{c}}{\bm{n}} $ and $  {\bm n}_{\rm tun}={\mathsf{P}}_{\mathrm{i}} {\bm{n}}$   with ${\bm{n}}$ being a generic vector with $N$ integer components, $\bm n=({n_1, n_2, \dots, n_N})^{\mathrm T}$.  The  integers in ${\bm{n}}$ can be of any sign  reflecting the fact that directions of backscattering (or tunneling) in continuous (or split)   channels can be opposite in different channels.

In the following section we will use the model formulated here for an RG analysis of the impact of the intra-channel local perturbation, \eqref{pert}, on the conductance of the multichannel LL.

\section{Scaling dimensions for scattering amplitudes \label{Sec3} }

The RG analysis of the impact of the scattering term, \eqref{pert}, requires the correlation functions of   the fields with the action defined by the Lagrangian density of \eqref{LN}.
Since the inter-channel interaction mixes the original channels, it is worth starting with the correlations in terms of the new fields, \eqref{M},  in which the Lagrangian of interacting multichannel LL is diagonal. To this end, we rewrite the boundary conditions of \eqref{bc} in terms of this fields:
\begin{align}\label{bc tilde}
  \Delta{\widetilde{\bm \theta}}(t)&=0\,, &\Delta{\widetilde{\bm \varphi}}(t) &=-2{\widetilde\Xi}\,{\widetilde{\bm \theta}}(0)\,,
 &{\widetilde\Xi}&= {\mathsf M}^{\rm T} \Xi \,{\mathsf M}.
\end{align}
This can be rewritten as in \eqref{in-out} via the   chiral fields, $\widetilde{\bm\varphi } _\mathrm{R,L} \equiv \widetilde{\bm\varphi }\pm \widetilde{\bm\theta }$, as
\begin{align}\label{S-tilde}
{\widetilde{ {\bm \Psi}}}_{\rm out}&= {{ {\mathsf{\widetilde{S}}}}}\,{ \widetilde{{\bm \Psi}}}_{\rm in},&\widetilde{\mathsf S}&=\begin{pmatrix}
                      \widetilde{{\mathsf{R}}} & \widetilde{{\mathsf{T}}} \\
                      \widetilde{{\mathsf{T}}} & \widetilde{{\mathsf{R}}} \\
                    \end{pmatrix},
\end{align}
where non-diagonal reflection and transmission matrices are related to $\widetilde{\Xi}$ by $ \widetilde{{\mathsf{T}}}=1-\widetilde{{\mathsf{R}}}=({1+\widetilde{\Xi}})^{-1} $, and ${\widetilde{ {\bm \Psi}}}_{\rm out}$ and ${\widetilde{ {\bm \Psi}}}_{\rm in}$  to $\widetilde{\bm\varphi } _\mathrm{R,L}({\pm0}) $  as    the original fields in \eqref{in-out}.

 The correlation functions of the fields $\widetilde{\bm\varphi} $ and $\widetilde{\bm \theta}$ with the Lagrangian density of \eqref{L tilde} can be easily found using  its diagonal form, \eqref{L-diag}. Incorporating   the above boundary conditions results in the following correlations
  of the local fields \cite{IVY:2013,*YY2014,KLY:16b} :
\begin{align}&\begin{aligned}
\langle{2 {\widetilde{{\bm\theta}} ({t})}} \otimes 2{\widetilde{\bm\theta }}^{\mathrm T} ({t'})\rangle&=-2\,\widetilde{{\mathsf{T}}}\,\ell\,, \\
\langle\Delta{\widetilde{\bm\varphi}}({t})\otimes\Delta{\widetilde{\bm\varphi}}^{\rm T} ({t'}) \rangle&=-2\,\widetilde{{\mathsf{R}}}\,\ell \,,
\end{aligned}\label{CFtilde} \end{align}
 where $\ell \equiv \ln(t-t') $. The  correlation functions of the original fields $\bm \theta $ and ${\bm{\varphi}}$ are obtained from the field transformation \eqref{M} as follows:
\begin{subequations}\label{CF} \begin{align} \label{CFa}
{ -\tfrac{1}{2}}\langle{2 { {{\bm\theta}} ({t})}} \otimes 2{ {\bm\theta }}^{\mathrm T} ({t'})\rangle&= {{\mathsf{M\widetilde{T}M^{\mathrm T}}}}\ell =[{{\mathsf{K}}^{-1}+\Xi }]^{-1}\ell ;  \\
{ -\tfrac{1}{2}}\langle\Delta{ {\bm\varphi}}({t})\otimes\Delta{ {\bm\varphi}}^{\rm T}({t'} )\rangle&= {{\mathsf{(M^{\mathrm T})^{-1} \widetilde{R}M^{-1} }}}\ell =[{{\mathsf{K}}+\Xi ^{-1}}]^{-1}\ell\,.
   \label{CFb}\end{align}\end{subequations}
Taking the physical limit described after \eqref{bc}  eliminates in \eqref{CFa} rows and columns corresponding to the continuous channels, and in \eqref{CFb}  rows and columns corresponding to the split channels. {The fact that the correlation functions are governed only by matrix ${\mathsf{K}}$    justifies referring to it as the  {Luttinger matrix}.}

The RG flow of each amplitude $v_{\bm n_{\mathrm{bs}},{\bm{n}_{\mathrm{tun}}}}$ describing different configurations of continuous and split channels in \eqref{pert} is defined by its scaling dimension, $ \Delta_{\mathrm{conf}} $. Using the correlation functions of \eqref{CF} to generalize the RG analysis \cite{KaneFis:92a} for the multichannel LL, we find  these dimensions as follows (see Appendix A for details):
  \begin{align}\label{Delta}
\Delta_{\mathrm{conf}}={\bm{n}}^{\mathrm T}  \left[{\mathsf{P}}_\mathrm{i}{\mathsf{K}}{\mathsf{P}}_\mathrm{i}+ {\mathsf{P}}_\mathrm{c}{\mathsf{K}}^{-1} {\mathsf{P}}_\mathrm{c}\right]^{-1}{\bm{n}}  \,.
  \end{align}
{The RG dimension $\Delta_{\mathrm{conf}}$  is fully governed by the Luttinger matrix ${\mathsf{K}}$, \eqref{K}.  Thus its role in defining the RG flows is similar to that of the Luttinger parameter $K$ for the single-channel LL. Any channel configuration  remains stable against embedding the local impurity, \eqref{pert}, as long as $\Delta_{\mathrm{conf}}>1$.
Obviously, $K_i=1$ does no longer separates the conducting and insulating state of the $i$-th channel. More interesting is that, generically, there exist regions in the phase diagram where both the conducting and insulating boundary conditions are either simultaneously stable or simultaneously unstable, as we detail in the following section for the two-channel LL. In the former case, a phase coexistence emerges where the parameters of the unperturbed Lagrangian \eqref{LN} do not determine the conducting state of the system: there should exist an unstable fixed point with the RG flows in its vicinity depending  on the scattering strength of the perturbation \eqref{pert}.  In the latter case, when neither zero nor ideal conductance is stable, it may flow  to an intermediate value  smaller than $e^2/h$, although there is no techniques, short of an exact solution, to determine this value. }

 \section{Two-channel liquid}\label{2chan}
 Here we   consider a two-channel LL implying that each channel has both right- and left-moving particles. In the absence of the inter-channel interaction such a two-channel LL has three distinct conducting configurations, as each of the two channels can be either conducting (labeled as `c')  or insulating (labeled as `i'). We analyze their RG stability with the interaction switched on. {The   RG dimension  in   \eqref{Delta} is fully governed by the three independent elements of the Luttinger matrix ${\mathsf{K}}$ that can be deduced from \eqref{LN}--(\ref{K1}). We start with some generic analysis in terms of the matrix elements of ${\mathsf{K}}$, and express these elements via the parameters of the Lagrangian in the subsequent section.}

\subsection{Generic analysis}\label{GenAn}
 The boundaries between different phases are governed by the stability conditions $\Delta_{{\mathrm{conf}}} >1$ , where  the RG dimension $\Delta_{{\mathrm{conf}}}$ is given by \eqref{2chConditions}, that must be satisfied for all the scattering processes (i.e.\ for $n_{1,2}=0,\pm1,\pm2\dots$). In this section we derive the parametric requirements for one- and two-particle scattering to dominate   \cite{KLY:16d}. For clarity, we explicitly rewrite the stability conditions  for all the two-channel configurations. We remind that  in the absence of the inter-channel interaction the  channels with $K_i>1$ ($K_j<1)$  remain  continuous (split) for any scattering strength.

\begin{enumerate} \begin{subequations}\label{2chConditions}
\item[{cc:}]  both channels are initially continuous. In this case the projectors in \eqref{Delta} are ${\mathsf{P}}_\mathrm{c}=\operatorname{diag}({1,1})$ and ${\mathsf{P}}_\mathrm{i}={\mathsf{0}}$, so that the  configuration is stable when
\begin{align}\label{cc}
\Delta_{\rm cc}=n_{1}^{2}K_{11}+2n_1n_2K_{12} +n_{2}^{2}K_{22}>1 \,.
\end{align}
 {\item[{ii:}] the channels are initially split,  ${\mathsf{P}}_\mathrm{i}=\operatorname{diag}({1,1})$ and ${\mathsf{P}}_\mathrm{c}={\mathsf{0}}$, so that the RG dimension is given by $\Delta_{\rm ii}=n_{1}^{2}({{\mathsf{K}}}^{-1}) _{11}+2n_1n_2({{\mathsf{K}}}^{-1})_{12} +n_{2}^{2}({{\mathsf{K}}}^{-1})_{22}$; expressing the elements of the inverse Luttinger matrix in terms of $\kappa\equiv\det {\mathsf{K}}>0$, we write the stability condition for this configuration as
    \begin{equation}\label{ii}
\kappa\Delta_{\rm ii}=n_{1}^{2}K_{22}-2n_1n_2K_{12} +n_{2}^{2}K_{11}>\kappa\,.
\end{equation}
 \item[{ic:}]    the first channel is initially continuous while the second is split,  ${\mathsf{P}}_\mathrm{i}=\diag({1,0})$ and ${\mathsf{P}}_\mathrm{c}=\diag({0,1})$; the configuration is stable when
\begin{equation}\label{ic}
\Delta_{\rm ic}=\frac{n_1^2}{ {K}_{11} }+\frac{\kappa n_2^2}{K_{11}} > 1 .
\end{equation}
\item[{ci:}]     here   ${\mathsf{P}}_\mathrm{i}=\diag({0,1})$ and ${\mathsf{P}}_\mathrm{c}=\diag({1,0})$ so that the stability condition is obtained by interchanging $1\rightleftarrows2$ in the r.h.s.\ of \eqref{ic}:
\begin{equation}\label{ci}
\Delta_{\rm ci}=\frac{n_2^2}{ {K}_{22} }+\frac{\kappa n_1^2}{K_{22}} > 1 .
\end{equation}}
\end{subequations}
\end{enumerate}
We will show in the next subsection that $K_{12} $  is proportional to the  inter-channel interaction strength. In its absence, when  $K_{12}=0 $, $K_{11}=K_1 $ and $K_{22}=K_2 $,  the following statements hold:  (i) the one-particle scattering is most relevant as the scaling dimensions in each channel are mutually independent; (ii) there is an obvious duality  \cite{KaneFis:92a} between WS and WL as $\Delta_{\mathrm{c}}=K $ and $\Delta_{\mathrm{i}}=1/K$ so that one (and only one) of the insulating or conducting phase is necessarily unstable.

None of these statements remains necessarily valid in the presence of the inter-channel interaction.  {We will show that the conditions in \eqref{cc} and \eqref{ii} can be simultaneously held in a certain parametric region, indicating the existence of an unstable critical point with RG flows being dependent on the scattering strength. Furthermore, for a sufficiently strong inter-channel interaction a multiple scattering becomes more RG relevant than the one-particle scattering  resulting in the conditions in \eqref{cc} and \eqref{ii} being simultaneously broken  \cite{KLY:16d}. }

Before illustrating this,  let us consider a straightforward  case  of no scattering in the conducting channel 2,  $n_2\equiv0$. This might happen when the channels are totally independent, e.g. they are spatially remote  or  have different physical nature, like in the electron-phonon LL  \cite{martin05,GYL:2011,*GYL:2011a,*YGYL:2013,IVY:2013,*YY2014}.  In this case channel 2 remains  conducting whereas one-particle scattering is  dominant in channel 1, so that for isolated channels are either in the cc (for $K_1>1$) or the ic (for $K_1<1$) configuration. The inter-channel    interaction shifts the boundary between the conducting and insulating behavior to $K_{11}=1 $  which now depends on characteristics of both channels. However, as $\Delta_{{\mathrm{cc}}}=K_{11}$ and $ \Delta_{{\mathrm{ic}}}=1/K_{11}$, the \emph{duality condition}, $\Delta_{{\mathrm{cc}}}\Delta_{{\mathrm{ic}}}=1   $, still holds.

 \begin{figure*}

 {\includegraphics[width=.32\textwidth]{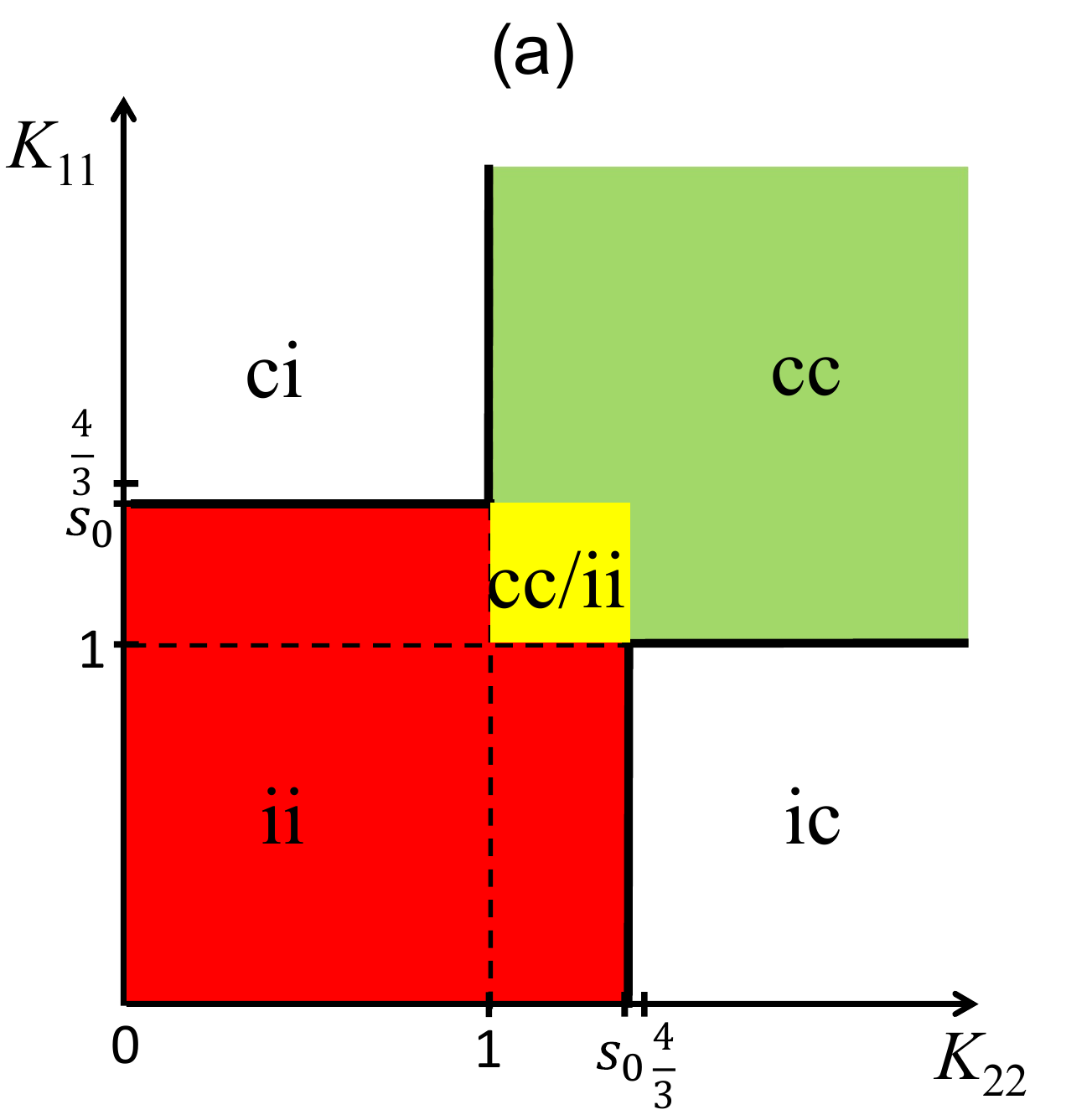}}\quad
 {\includegraphics[width=.32\textwidth]{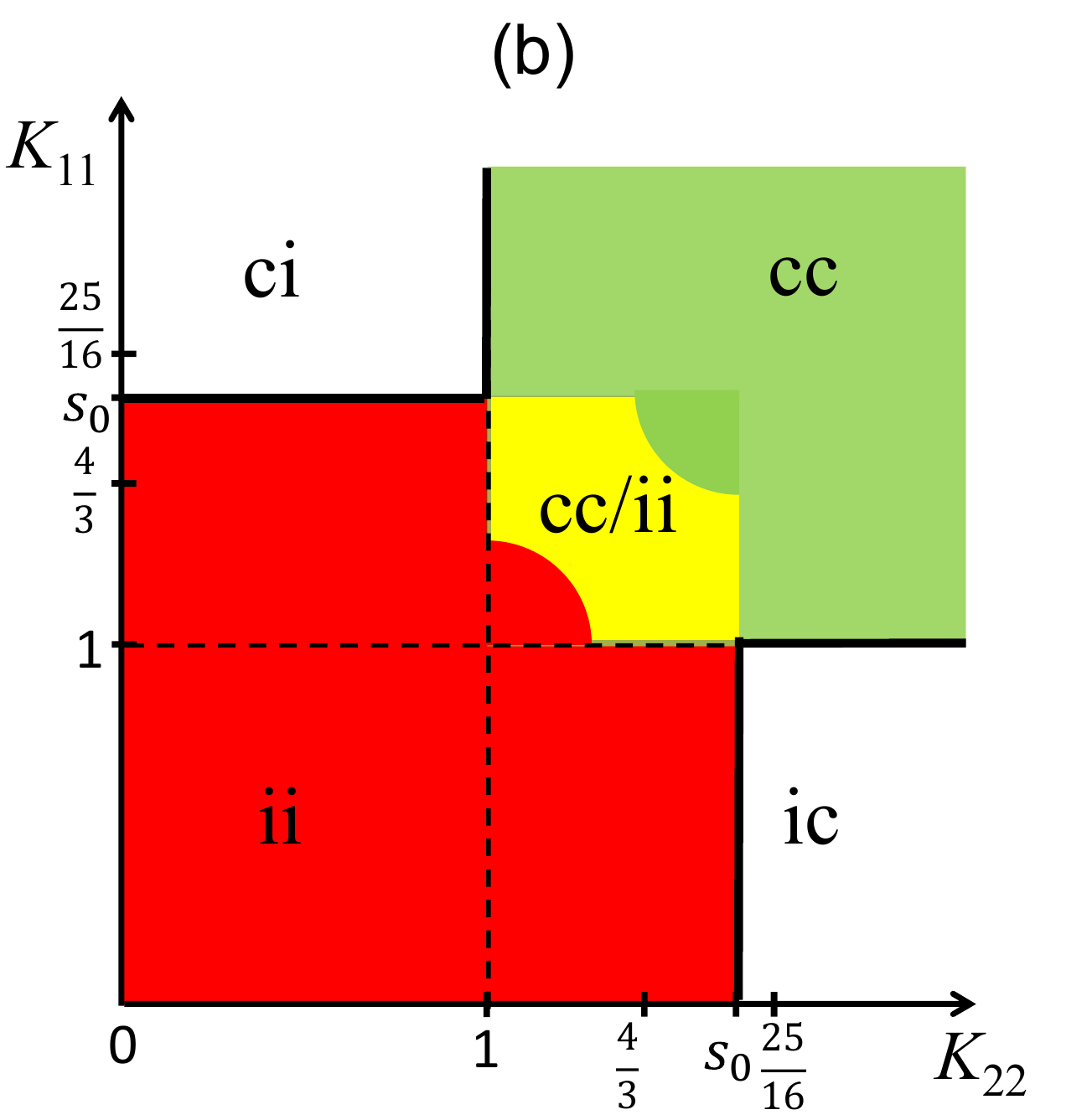}}\quad
 {\includegraphics[width=.32\textwidth]{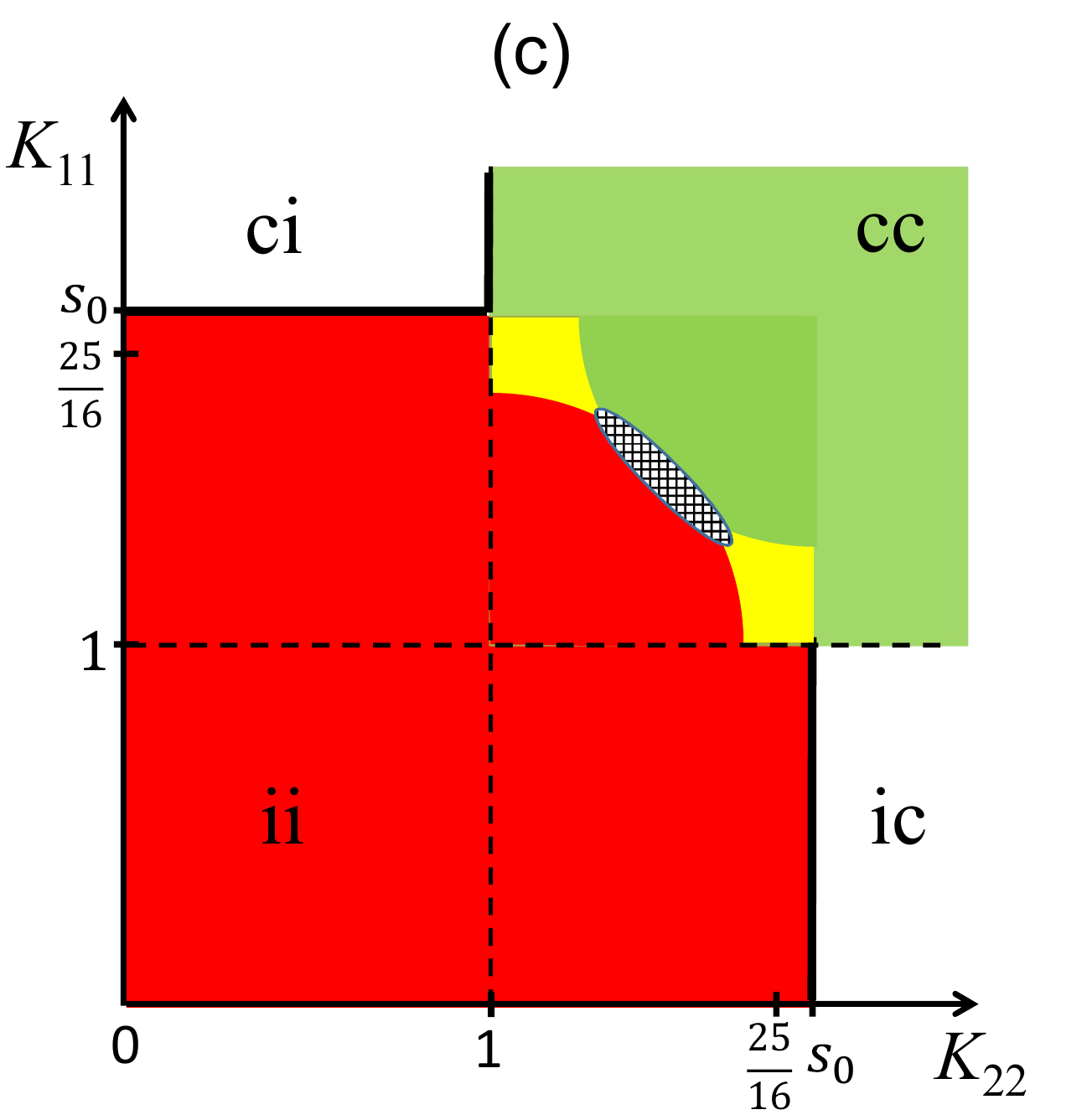}  }

\caption{Schematic phase diagrams: (a) Under condition \eqref{SufCond}, when  the one-particle scattering  is dominant for all $K_{11} $ and $K_{22} $, the ${\mathrm{cc}}$    and   ${\mathrm{ii}}$ phases   are both stable with respect to one-particle scattering from a WS or a WL, respectively, in the square $1<K_{11},\, K_{22} < s_0$.\; (b)  With the inter-channel interaction   increasing,   the square of the ${\mathrm{cc}}$ - ${\mathrm{ii}}$ phase  coexistence  grows; however, for $s_0>\frac{4}{3}$ the ${\mathrm{cc}}$ (or {${\mathrm{ii}}$}) phase becomes unstable with respect to two-particle scattering from a WS (or WL) in the lower (upper) corner of this square.  (c) With further increase of $K_{12}$, at $|\cos \gamma|>3/5$, the two-particle scattering results in the appearance of the cross-hatched region on the phase diagram where both the ${\mathrm{ii}} $ and ${\mathrm{cc}}$ phases are unstable. }
\label{simple}
\end{figure*}

When scattering is possible in both channels, a more complicated picture emerges. To analyze which scattering configuration is RG dominant we represent ${\mathsf{K}}$ {as a Gram matrix built of two vectors}, $\{{K_{ij}}\}=\bm g_i\cdot \bm g_j $, where $|{\bm{g}}_i|=\sqrt{K_{ii} }$, while the angle $\gamma=\widehat{{\bm{g}}_1{\bm{g}}_2}$ is given by
\begin{align}\label{gamma}
    \cos \gamma=\frac{K_{12}}{\sqrt{K_{11} K_{22}  } }
\end{align}
 {Such a representation is possible when the inter-channel interaction is not too strong: for $K_{12}\geqslant\sqrt{K_{11}K_{22}  }$ one enters the region of the Wentzel--Bardeen instability \cite{KLY:16a} where the channels should be totally restructured.  In the subsequent analysis we will stay clear of this region. In this representation} $\Delta_{\mathrm{cc}}={\bm{G}}^2$ where ${\bm{G}}=n_1{\bm{g}}_1+n_2{\bm{g}}_2, $ and a similar expression holds for $\Delta_{{\mathrm{ii}}} $ in terms of the inverse Luttinger matrix. Then the problem of finding a configuration corresponding to the most RG relevant scattering (which has the smallest $\Delta_{\mathrm{conf}}$) is reduced to that of finding the shortest vector on a $2D$ lattice spanned by ${\bm{g}}_1$ and ${\bm{g}}_2$. In general, this shortest vector problem (SVP) does not have an analytic solution and is known to be computationally hard   \cite{SVP:1,*SVP}.  It is, however, possible to formulate the parametric conditions for which  one-particle scattering, $n_1=1,\,n_2=0$ or $n_1=0,\,n_2=1$, dominates the RG flows  \cite{KLY:16d}.  It is shown in \emph{Appendix B} that the \emph{sufficient} condition for one-particle scattering to dominate is
 \begin{align}\label{SufCond}
    K_{12}<\tfrac{1}{2}\min\{{K_{11},\,K_{22}  }\}\quad \Leftrightarrow \quad |\cos\gamma|<\tfrac{1}{2}.
 \end{align}
 As $K_{12} $  is proportional to the  inter-channel interaction strength, the above inequality  holds when this interaction is  sufficiently small.

To determine the boundaries between nontrivial phases in this case, we substitute $|n_1|=1, \, n_2=0$ or $n_1=0$, $  |n_2|=1$ into the stability conditions of \eqref{2chConditions}. Expressing $\kappa\equiv\det {\mathsf{K}}$ in terms of $\gamma$ as $\kappa=K_{11}K_{22}\sin ^2\gamma  $, we represent these conditions (with $s_0\equiv  1/ {\sin^2\gamma}$) as
\begin{align}\label{2ch1pcond}
   \begin{aligned} K_{11},\,K_{22}&>1 \;\;({\mathrm{cc}});&K_{11},\,K_{22}&<s_0  &&(\mathrm{ii});\\
      \!\!\!K_{11}<1,\,K_{22}&>s_0\; ({\mathrm{ic}});& K_{22}<1,\, K_{11}&> s_0 &&(\mathrm{ci}).
\end{aligned}\end{align}
Since $s_0>1$ the boundaries of the cc and ii phases inevitably overlap, as illustrated in Fig.~\ref{simple}(a): inside the central square, i.e.\ for $1<K_{11},\,K_{22}<1/\sin^2\gamma  $, the phase where both channels are conducting is stable against weak scattering while the phase where both are insulating is stable against weak tunneling. As the elements of ${\mathsf{K}}$ are the same for both phases, they can be only distinguished by the impurity scattering strength implicit in \eqref{pert}. Therefore, a new unstable fixed point characterized by some critical value of scattering should exist for any given ${\mathsf{K}}$. Such a scattering-dependent fixed point describes a  transition between insulating and conducting phases simultaneous for both channels. Any transition between the ${\mathrm{c}}$ and ${\mathrm{i}}$ phases that happens only in one of the channels is fully defined by the parameters of the Lagrangian in \eqref{LN} independently of the scattering strength. This is illustrated by the solid phase boundaries between ${\mathrm{ii}}$ and ${\mathrm{ic}}$ phases, etc., in Fig.~\ref{simple}.

When the inequality in \eqref{SufCond} fails  with increasing $K_{12} $, \eqref{gamma}, which characterizes the inter-channel interaction,  the  one-particle scattering  still dominates in certain parts of the phase diagram; the appropriate necessary  conditions  are  derived in Appendix B.  However,  many-particle (first of all, two-particle) scattering starts to change the phase diagram.  Note that the change affects only the ${\mathrm{cc}}$ and ${\mathrm{ii}}$ phases while the stability of the ${\mathrm{ic}}$ or ${\mathrm{ci}}$ phases is unaffected by the many-particle scattering, as seen from Eq.~(\ref{2chConditions}c,d).

We consider  the most relevant case \cite{KLY:16d} of the stability conditions, Eq.~(\ref{2chConditions}a,b), broken  by the two-particle scattering.    Substituting $|n_{1,2}|=1$ into   \eqref{cc} and \eqref{ii}  we find  the two-particle \emph{instability} conditions as follows:
\begin{align}\label{2p-unstable}
    \begin{aligned}\!\!\!\!K_{11}\pm 2\sqrt{K_{11}K_{22}} \cos \gamma\!+\!K_{22} &<1;   &(\mathrm{cc})&\\
    \!\!\!\!K_{11}\pm 2\sqrt{K_{11}K_{22}} \cos \gamma\!+\!K_{22}&<  K_{11}K_{22}\sin^2\gamma. \! \!\!\! &(\mathrm{ii})&\end{aligned}
\end{align}
When $|\cos \gamma|>\frac{1}{2}$ (i.e.\ $s_0>\frac{4}{3}$), both these inequalities hold inside the parametric region of \eqref{2ch1pcond}, where the $\mathrm{ii}$ and $\mathrm{cc}$ phases are \emph{stable}  with respect to the one-particle scattering.
Thus going beyond the one-particle stability condition, \eqref{SufCond}, results in a more complicated form of  the region of the phase coexistence  as schematically illustrated in Fig.~\ref{simple}(b).  There the yellow square,  corresponding to the region of simultaneous stability of the $\mathrm{ii}$ and $\mathrm{cc}$ phases with respect to the one-particle scattering,  increases; however, both these phases become unstable with respect to two-particle scattering in the corners of this square.  The  exact shape of the phase boundaries is {not relevant but can be easily found from \eqref{2p-unstable}}.

\begin{figure*}
{\includegraphics[width=.3\textwidth]{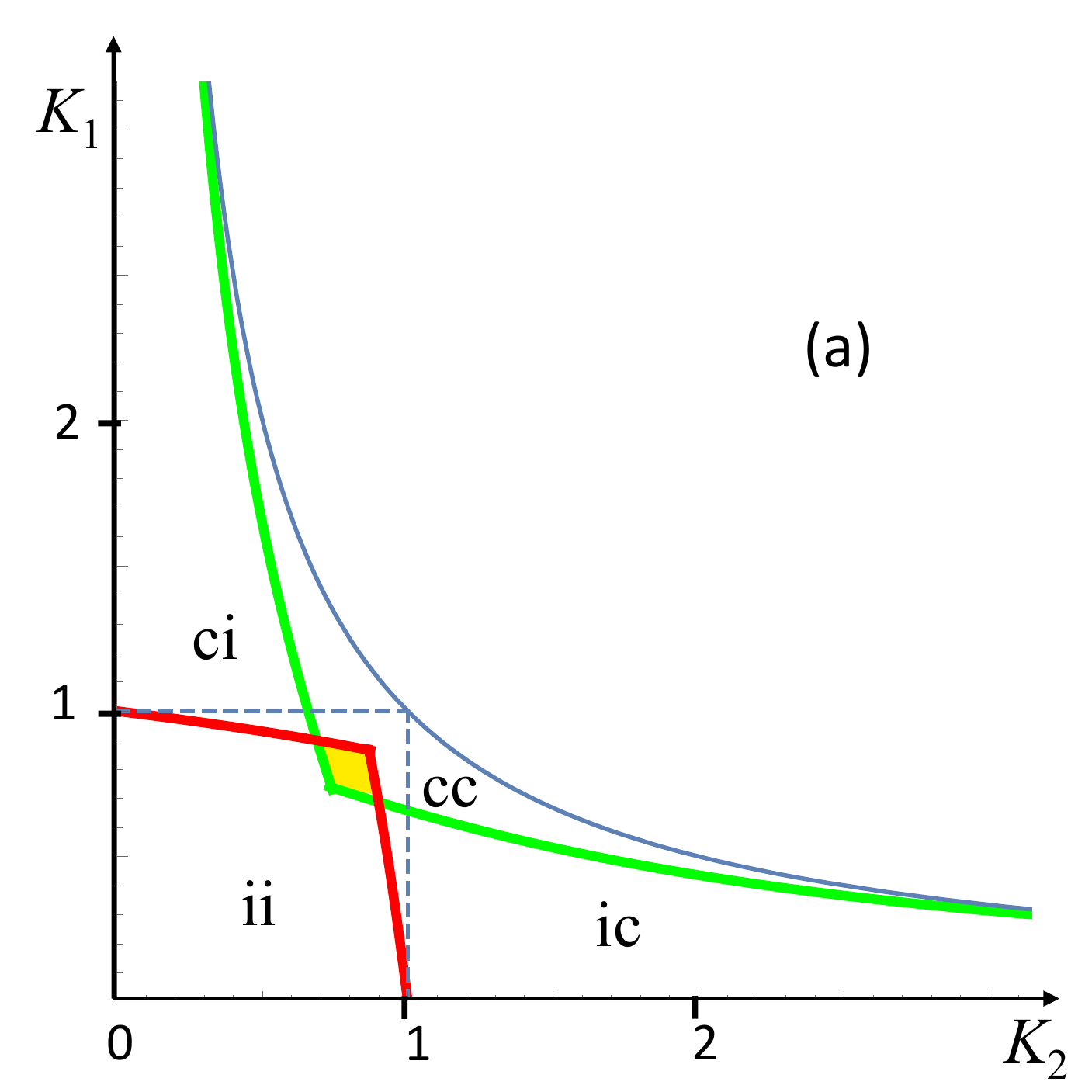}  }\qquad
{\includegraphics[width=.3\textwidth]{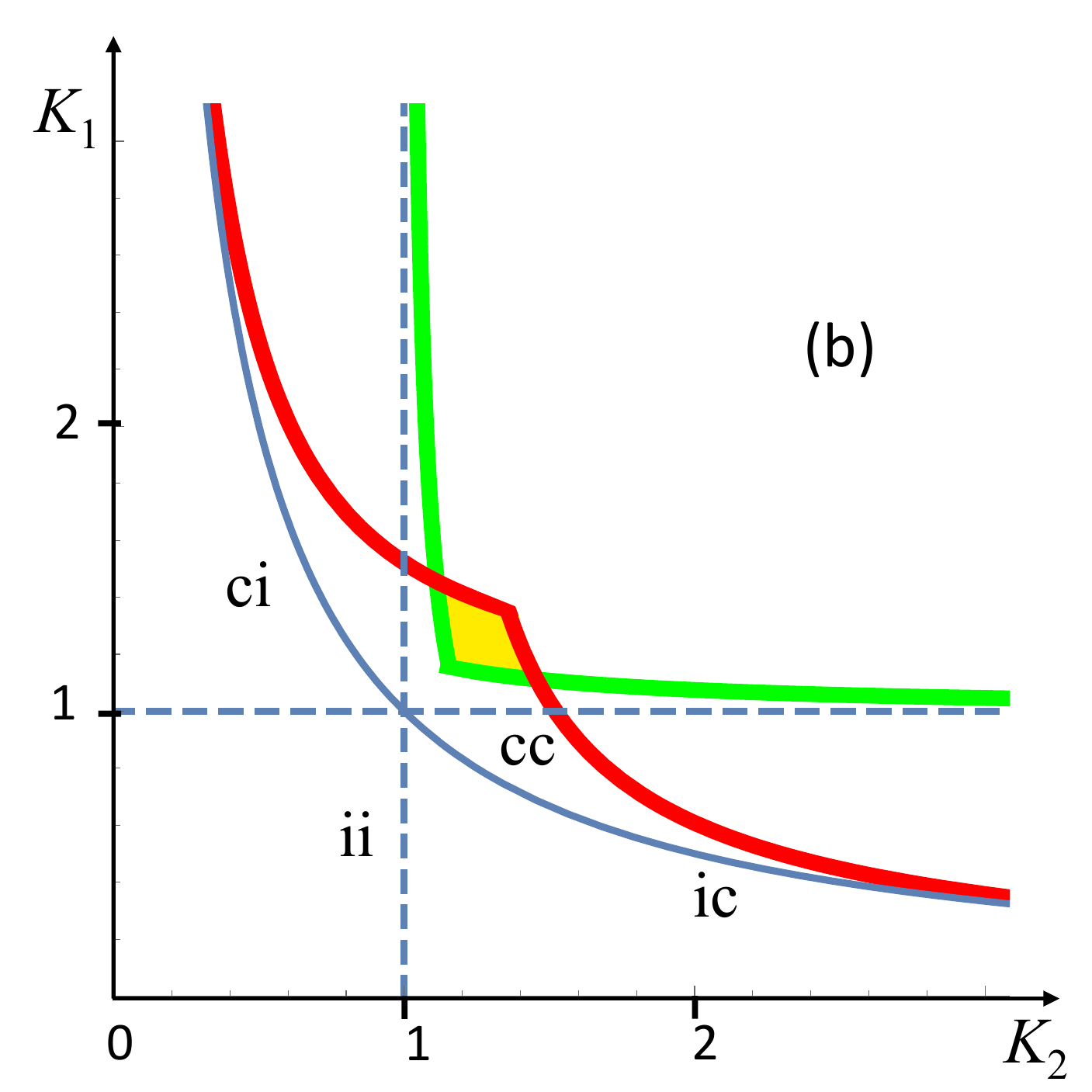}  }\qquad
{\includegraphics[width=.3\textwidth]{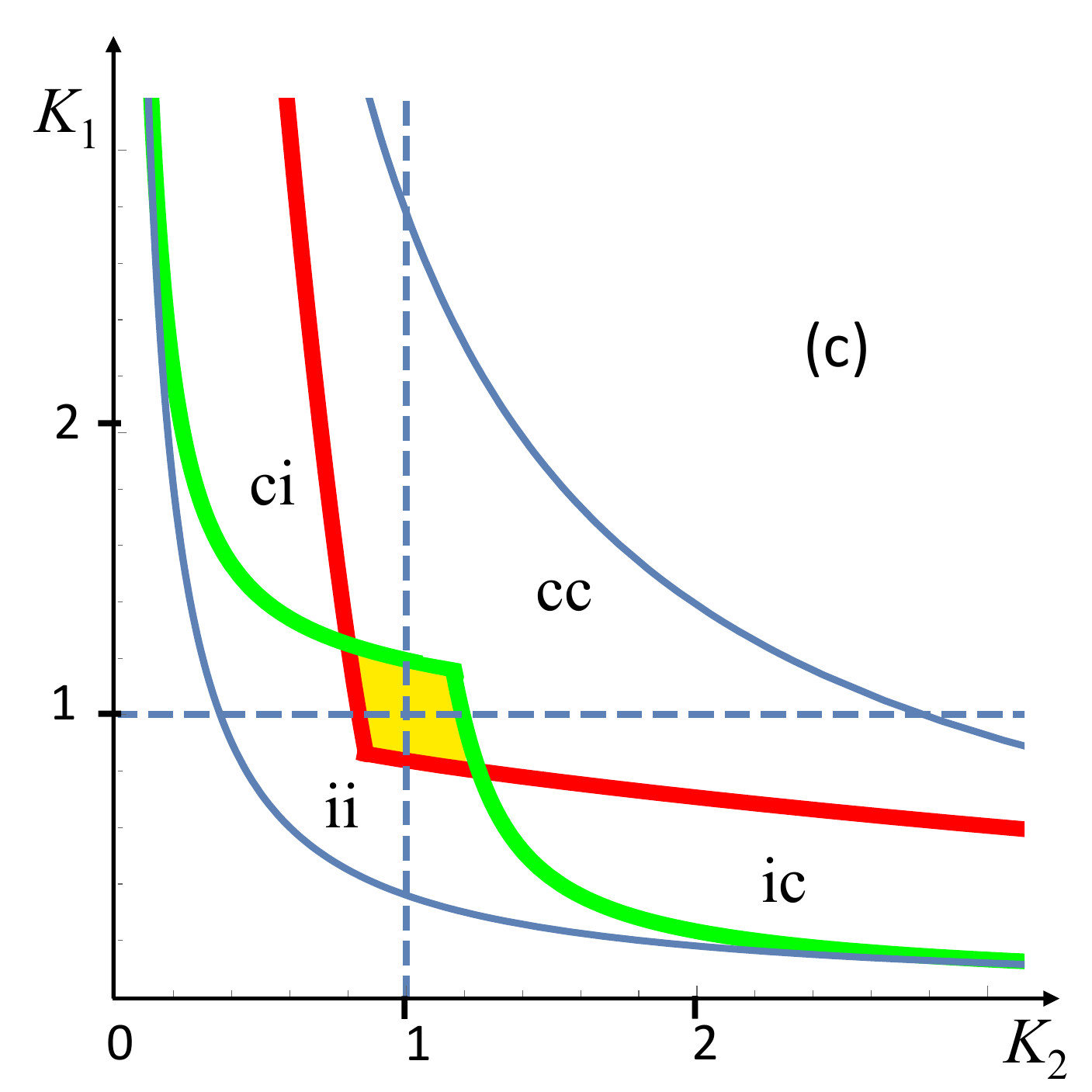}  }
\caption{Phase diagrams for $\beta=1$ and different values of the inter-channel interaction parameters:
(a) ${\widetilde{g}}_4=\widetilde{g}_2=0.5\,$;  (b) $\widetilde{g}_4=-\widetilde{g}_2=0.5\,$; (c) $\widetilde{g}_4=0;\;\widetilde{g}_2=0.6\,$. The blue curves on each graph show the boundary of the Wentzel -- Bardeen instability region,  \cite{KLY:16a} while in the yellow region both ${\mathrm{ii}}$ and ${\mathrm{cc}}$ phases are stable with respect to one-particle scattering. {{Here we assume that the condition of \eqref{SufCond} is fulfilled, i.e.\ multiple scattering is irrelevant.}}}
\label{exact}
\end{figure*}

With  further increase of the inter-channel interaction, \eqref{gamma},  the regions of the two-particle instability start to overlap when both the inequalities in \eqref{2p-unstable} hold simultaneously, see Fig.~\ref{simple}(c). This first happens in the center of the phase coexistence region, where $K_{11}=K_{22}  $ which gives $|\cos\gamma|=\frac{3}{5}$ (i.e.\ $s_0=\frac{25}{16}$). Thus a totally new situation might emerge \cite{KLY:16c}   for $|\cos\gamma|>\frac{3}{5}$  where the ${\mathrm{cc}}$ phase is unstable against weak scattering, while the ${\mathrm{ii}}$ phase is unstable against a weak link. This signals the existence of a non-trivial \emph{attractive} fixed point at some intermediate value of the scattering strength. Again, the RG flows in its vicinity depend on the impurity scattering strength.  The conductance of each channel in such a case is finite, but smaller than the ideal value. It might be possible in such a case to redefine the channels so that one of them would become fully insulating while the other ideally conducting, as we illustrate in Section \ref{topo}.

 \subsection{Scattering boundaries in two-channel LL}

The elements of the Luttinger matrix that define the phase stability conditions and thus the boundaries of all the phases are implicitly dependent on the inter-channel interaction strengths, $V_{\theta,\varphi } $, as well as on the particle velocities, $v_{1,2} $, and the Luttinger parameters, $K_{1,2} $, in both the channels. Here we explicitly derive this dependence.

 The $2\times 2$ Luttinger matrix ${\mathsf{K}}$, which governs the stability conditions \eqref{2chConditions}, is defined via the interaction matrices, ${\mathsf{V}}_{\theta}$ and ${\mathsf{V}}_{\varphi}$, by \eqref{K}.    Now we will express ${\mathsf{K}}$ explicitly in terms of matrix elements of ${\mathsf{V}}_{\theta,\varphi}$.  In the two-channel case these  matrices, which define the Lagrangian \eqref{LN},   {are represented in terms of  the inter-channel density-density and current-current interaction strengths,  $V_\varphi $ and $V_\theta $,  the Luttinger parameters, $K_{1,2} $, and renormalized velocities, $v_{1,2} $, in each channel as follows}:
 \begin{align}\label{VV}
 {\mathsf{V}} _\theta &=
  \begin{pmatrix}
  v_1 K_1^{- 1} & V_{\theta} \\
  V_{\theta} & v_2 K_2^{- 1} \\
  \end{pmatrix}
,
&
 {\mathsf{V}} _{\varphi}&=
     \begin{pmatrix}
  v_1 K_1 & V_{\varphi} \\
  V_{\varphi} & v_2 K_2 \\
   \end{pmatrix}.
\end{align}
 Using the fact that the determinants of ${\mathsf{V}}_{{\theta,\varphi } } $ are positive  \cite{KLY:16a}, we represent them as
 \begin{subequations}\label{alpha}\begin{align}
\det{\mathsf{V}}_\theta &=  \frac{v_1v_2 \cos^2\alpha_\theta}{K_1K_2} ,&
\det{\mathsf{V}}_\varphi&= {v_1v_2K_1K_2}\cos^2\alpha_\varphi,
\end{align}
where
{\begin{align}\begin{aligned}
        \sin \alpha_\theta &\equiv \sqrt{\frac{K_1K_2}{v_1v_2}}V_\theta
        \equiv\sqrt{K_1K_2} ({\widetilde{g}_{4 }+ \widetilde{g}_{2}}), \\
 \sin \alpha_\varphi&\equiv \frac{V_\varphi }{\sqrt{  K_1K_2}} \equiv
 \frac{{\widetilde{g}_{4 }-\widetilde{g}_{2}}}{\sqrt{  K_1K_2}} ,
\end{aligned}\end{align}}\end{subequations}
and $\widetilde{g}_{4,2}$ characterize inter-channel interactions of the density components of the same or opposite chirality.
Substituting this into \eqref{K1}, with ${\mathsf{A}}\equiv{\mathsf{V}}_\theta $ and ${\mathsf{B}}\equiv{\mathsf{V}}_\varphi $, we arrive at the following representation of the Luttinger matrix:
 \begin{align}\notag
   {\mathsf{K}}
    &=
                                \frac{1}{B }
                                \begin{pmatrix}
                                   K_1(\beta +  \rho) & {\frac{  \sqrt{\beta K_1K_2} \sin({\alpha_\varphi- \alpha_\theta })}{\cos \alpha_\theta }} \\[6pt]
                                        {\frac{  \sqrt{\beta K_1K_2} \sin({\alpha_\varphi- \alpha_\theta })}{ \cos \alpha_\theta}}
                                         & K_2({1+\beta\rho})
                                   \end{pmatrix} \\[6pt]\label{KK}
                                   &=\begin{pmatrix}
                                            {K_1}/{K_{{\mathrm{c}}1} } & K_{12}  \\[6pt]
                                            K_{12}  &    {K_2}/{K_{{\mathrm{c}}2} } \\
                                          \end{pmatrix}, \quad\begin{array}{lcl}
                                                           K_{\mathrm{c}1}&\equiv&\frac{B}{\beta+\rho} \\[6pt]
                                                           K _{\mathrm{c}2}&\equiv&\frac{B}{1+\beta\rho}
                                                         \end{array}
\end{align}
where $
  B \equiv \sqrt{1+\beta^2+2 \beta  \cos({\alpha_\varphi- \alpha_\theta })},
$ $
    \beta \equiv {v_1} /{v_2}  , $ and $\rho \equiv{\cos\alpha_\varphi }/{\cos \alpha_\theta} .$

To express the phase  boundaries     in Fig.~\ref{simple} in these terms, we note that $\kappa\equiv\det {\mathsf{K}}=K_1K_2|\rho|$ as follows from \eqref{K} and \eqref{alpha}.
 On the other hand, $\kappa=K_{11}K_{22}\sin^2\gamma  $. Therefore, substituting
 $1/s_0\equiv \sin ^2\gamma = \rho K_{{\mathrm{c1}}}K_{\mathrm{c2}}$ in \eqref{2ch1pcond} we find
\begin{align}\label{Boundaries}
&\begin{aligned}
    K_{1}&>K_{{\mathrm{c}}1}  \cr
    K_2&>K_{\mathrm{c}2}
\end{aligned}
 {\mathrm{(cc)}}  & \begin{aligned} K_1&<1/\rho{K_{{\mathrm{c}}2} };  \cr  K_2 &< 1/ \rho{K_{{\mathrm{c}}1} } \end{aligned} {\mathrm{(ii)}}.
\end{align}
{Thus the phase diagram with allowance only for the one-particle scattering looks on the \mbox{$K_1$ - $K_2$} plane exactly as that in Fig.~\ref{simple}(a) with the straight boundaries being defined by the   inequalities \eqref{Boundaries}.}

However, such a picture is deceptively simple:  both $K_{{\mathrm{c}}1 }   $ and $K_{{\mathrm{c}}2} $  nontrivially depend  on   the five parameters in \eqref{KK} that define the clean two-channel Luttinger liquid: the Luttinger parameters in each channel themselves, the velocity ratio, and the two inter-channel interaction parameters.   We illustrate such a dependence by fixing the values of some of these parameters. Choosing $v_1=v_2\equiv v$ simplifies the expressions for the boundaries: it follows from  \eqref{KK} that $K_\mathrm{c}\equiv K_{\mathrm{c1}}=K_\mathrm{c2}= \cos \alpha_\theta/\cos \frac{1}{2}({\alpha_\theta+\alpha_\varphi })$ and $1/\rho K_\mathrm{c}= \cos \frac{1}{2}({\alpha_\theta+\alpha_\varphi })/\cos\alpha_\varphi $. Specifying three different choices of the inter-channel interaction in \eqref{alpha} via $\widetilde{g}_{4,2}$, with $V_{\theta,\varphi }\equiv v({\widetilde{g}_{4 }\pm \widetilde{g}_{2}}) $, we arrive at three examples in Fig.~\ref{exact}. Note that, although we have chosen $\beta=1$ for illustrations, there is an important robust  feature on these phase diagram: for any $\beta$ the yellow region, representing the $\mathrm{cc}$-$\mathrm{ii}$ phase coexistence, is always  below the lines $K_{1,2}=1$ for $\alpha_\theta>\alpha_\varphi>0 $ (a), or above these lines for $\alpha_\varphi>\alpha_\theta>0$ (b), while the noninteracting point $K_1=K_2=1$ is inside these region when the signs of the inter-channel interaction parameters $\alpha_{\theta,\varphi } $ are opposite (c). We do not show in Fig.~\ref{exact} the boundaries of two-particle instability, which is analytically obtained   in \emph{Appendix B}   by substituting elements of matrix ${\mathsf{K}}$, \eqref{KK}, into   condition \eqref{2p-unstable}.

\section{\label{topo}Weak scatterer and weak link in two-channel topological insulators}
{Now let us consider in more detail another example of a two-channel LL: a 2D topological insulator supporting two helical states at each edge  \cite{PRL09-Chamon, Santos2015}.  We analyze whether  current-carrying edge states remain stable against potential scattering as in \eqref{pert}. The time-reversal symmetry forbids intra-edge scattering, while a spin-conserving backscattering between the edges is allowed. The scattering amplitude can be regulated by a distance between the edge states and can be locally increased when they approach each other, e.g., like in quantum point contacts (QPC) in a narrow Hall bar geometry \cite{Heiblum:15}.
Assuming that both  these channels  are of the same physical nature so that $K_1=K_2\equiv K$ and $\beta\equiv v_1/v_2=1$, it is convenient to form the initial channels  from  spin-up and spin-down electrons so that left- and right-movers in each channel belong to the opposite edge.  The backscattering then becomes an intra-channel process while the inter-channel scattering is forbidden by time-reversal symmetry.

With such a choice {of the channels, the present case} falls within the generic analysis of the previous sections. The  two-channel Luttinger matrix \eqref{KK} simplifies:
\begin{align}\label{K3}
    {\mathsf{K}}&=  \dfrac{K{\mathrm{sgn} [\cos\frac{\alpha_\varphi -\alpha_\theta}{2}  ]}}{ \cos\alpha_\theta}\begin{pmatrix}
                                                   \cos\frac{\alpha_\varphi +\alpha_\theta}{2} & \sin   \frac{\alpha_\varphi -\alpha_\theta}{2}  \\[4pt]
                                                   \sin   \frac{\alpha_\varphi -\alpha_\theta}{2}   &   \cos\frac{\alpha_\varphi +\alpha_\theta}{2}  \\
                                                 \end{pmatrix}.
 \end{align}
Note that in this case the mixed ${\mathrm{ci/ic}}$ phases are inevitably unstable against   one-particle scattering as the diagonal elements of the Luttinger matrix above are equal to each other thus violating  the stability conditions for these phases in \eqref{2ch1pcond}.

\begin{figure}[b]\centerline{ {\includegraphics[width=.45\textwidth]{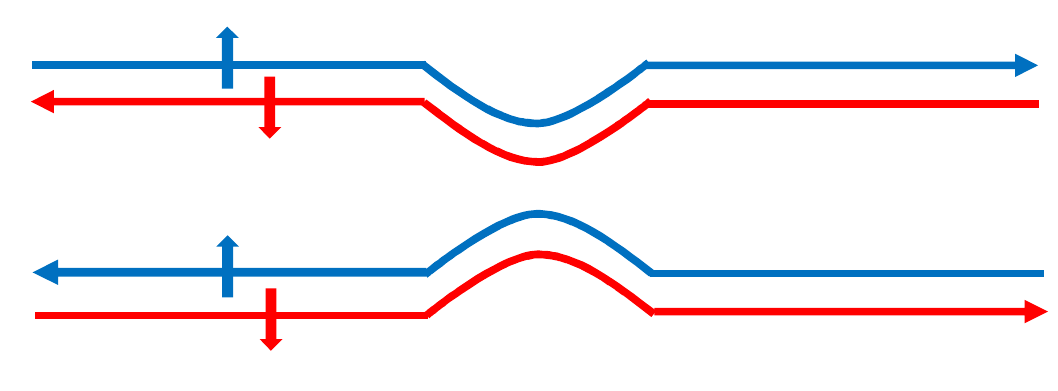}}}
\caption{\label{Fig3}{{Helical edge currents in a topological insulator with QPC. We re-label the channels so that spin-up electrons at the opposite edges form one channel and spin-down the other. In this nomenclature, only a local intra-channel scattering is allowed at the QPC since the inter-channel one is forbidden by the time-reversal symmetry. On the other hand,   the interaction between the modes of opposite helicity at each edge is translated into the inter-channel interaction while the intra-channel one is suppressed due to a spatial separation between the modes belonging to the same channel. }}}
\end{figure}
It is reasonable to assume that only particles at the same edge are interacting (apart from relatively short regions of QPC where the interaction can be absorbed into the scattering coefficients). Then the inter-channel interaction is always between the particles of the opposite chirality,  $\widetilde{g}_4=0$, i.e.\ $V_{\theta}=-V_\varphi  $ in \eqref{alpha} resulting in  $\sin \alpha_\varphi =-\sin \alpha_\theta/K^2$ so that the Luttinger matrix \eqref{K3} depends only on two parameters.
{In a particular case of the channels in Fig.~\ref{Fig3}  built} from the interacting electrons, the intra-channel interaction contains only $g_4$-proportional term  resulting \cite{Giamarchi} in $K=1$.
 In this case   from \mbox{$V_\theta=-V_\varphi $} follows $\alpha_\theta=-\alpha_\varphi \equiv\alpha$ so that \eqref{K3} reduces to
 \begin{align}\label{K4}
    {\mathsf{K}}&=   \frac{1}{|\cos \alpha|}  \begin{pmatrix}
       1 & -\sin \alpha \\
       -\sin \alpha & 1 \\
     \end{pmatrix}.
 \end{align}
Graphically, this state  corresponds to the middle point, $K_1=K_2=1$, in the phase diagram (c) in Fig.~\ref{exact} which lies in the ${\mathrm{ii}}$ -- ${\mathrm{cc}}$ phase coexistence region  where both these phases are stable  with respect to one-particle scattering.  There the ultimate choice of the phase depends on the impurity scattering strength. Thus, although the ${\mathrm{cc}}$ state is protected against weak scattering,  as has been noted earlier  \cite{Santos2015},   no protection against strong  scattering exists.

Even such a limited `protection' fails with increasing the inter-channel interaction so that  two-particle scattering becomes relevant. This happens at $|\sin \alpha|\geqslant\frac{3}{5}$ after the two instability regions meet at the center, $K_1=K_2=1$,  in Fig.~\ref{simple}(c). To prove this, it is worth rewriting the RG exponents for the ${\mathrm{ii}}$ and ${\mathrm{cc}}$ phases, \eqref{2chConditions}, for the present case:
  \begin{align}\label{mix}
    \Delta_{{\mathrm{cc}/\mathrm{ii}}} =\frac{n_\uparrow^2+n_\downarrow^2\mp 2 n_\uparrow n _\downarrow\sin\alpha}{|\cos\alpha|} . \end{align}
 For the two-particle scattering, $|n_\uparrow|=|n_\downarrow|=1$, these exponents are  smaller than $1$ (making the phases unstable) for $|\sin \alpha|>\frac{3}{5}$. Naturally, this condition is equivalent to the general one, $|\cos \gamma|>\frac{3}{5} $, Fig.~\ref{simple}(c), as it follows from the definition of $\gamma$, \eqref{gamma}, that  $ \cos\gamma  =-\sin \alpha $ for the matrix \eqref{K4}.
 Under this condition there should exist, as described earlier,  an intermediate  \emph{stable} fixed point corresponding to finite conductance of both the spin-up and spin-down channel.

 Such a finite conductance, however, usually signifies the possibility of introducing composite channels, one continuous (ideal conductance) and one split (no conductance). In the present case, they correspond to the standard `charge-spin separation' choice of channels.   Indeed, introducing $n_{\mathrm{ch}}= n_\uparrow+n_\downarrow $ and  $n_\mathrm{sp}= n_\uparrow-n_\downarrow $ diagonalizes \eqref{mix} for the RG exponents: $\Delta_\mathrm{cc}=\frac{1}{2}(K_\mathrm{ch}n_\mathrm{ch}^2+K_\mathrm{sp}n_\mathrm{sp}^2)$ and $\Delta_\mathrm{ii}=\frac{1}{2}(K_\mathrm{ch}^{-1} n_\mathrm{ch}^2+K_\mathrm{sp}^{-1} n_\mathrm{sp}^2)$
where $K_\mathrm{ch}= K_\mathrm{sp}^{-1}= ({1-\sin\alpha})/|\cos\alpha| $.  As $n_\mathrm{ch}+n_\mathrm{sp}$ must be even, the lowest order scattering process is $|n_\mathrm{ch}|=|n_\mathrm{sp}|=1$, corresponding to the (RG irrelevant) one-particle scattering in the `old' spin-up and spin-down channels. For such a process $\Delta_{\mathrm{cc}}=\Delta_{\mathrm{ii}}=\frac{1}{2}({K_{\mathrm{ch}}+ K_{\mathrm{ch}}^{-1} })=1/|\cos\alpha|>1$.

 The lowest-order  charge-only ({$|n_\mathrm{ch}|=2$}) or spin-only  ({$|n_\mathrm{sp}|=2$}) scattering processes correspond to the two-particle scattering in the  `old' channels with $n_\uparrow=n_\downarrow=\pm1$ or $n_\uparrow=-n_\downarrow=\pm1$, respectively.  Thus, although both the  ${\mathrm{cc}}$ and  ${\mathrm{ii}}$ phases are unstable with respect to the two-particle scattering for $|\sin \alpha|>\frac{3}{5}$, the instability reveals itself in different ways depending on the sign of $\alpha$. For the repulsive inter-channel interaction ($\alpha>0$),  the  charge channel becomes insulating while  the spin one remains ideally conducting, while for the attractive interaction ( $\alpha<0$) the roles of the charge and spin channels are inverted.  For the weak or intermediate inter-channel interaction, $|\sin \alpha|<\frac{3}{5}$, both new channels remain conducting so that both ${\mathrm{cc}}$ and ${\mathrm{ii}}$ phases remain stable, corresponding to the existence of an unstable fixed point with RG flows depending on the scattering strength.

 Any two-channel LL  with  the intra-channel interaction and inter-channel scattering   suppressed fits into the scenario described in this section. In particular, it  reproduces the earlier result \cite{PRL09-Chamon} on a corner junction  between the edge currents in topological insulators.  Let us also repeat that the idea of `interaction-protected' transport verified for weak scattering \cite{Santos2015}  needs analysis also for strong scattering (weak links). The results of this section show that  for any intra-level interaction the edge currents are only stable against weak scattering, while allowing for two-particle scattering  in the presence of a sufficiently strong intra-level interaction completely suppresses the edge currents.

\section{Conclusion}
We have developed a powerful approach to deal with a local impurity in multichannel Luttinger liquids. We have identified the Luttinger matrix,  \eqref{d2}, (\ref{K}) and (\ref{KK}), that controls   scaling dimensions of all perturbations in all possible phases. {Thus we have obtained the phase diagram for a generic
two-channel Luttinger liquid, Fig.~\ref{simple}, that in certain parametric regions is governed by multiple  scattering  from the impurity \cite{KLY:16d}.  We have constructed the phase boundaries that depend on the strength of inter-channel interaction as well as on the intra-channel LL characteristics, Fig.~\ref{exact}. The presented approach is applicable to channels of different nature as in fermion-boson mixtures, or to identical ones as on the opposite edges of a topological insulator. In the future we will extend it to particular interesting cases of a multi-channel LL. }

\section*{Acknowledgments}
IVY research was funded by the Leverhulme Trust Research Project Grant RPG-2016-044.

\begin{appendix}
    \section{Scaling dimensions}
As the Lagrangian in terms of the fields $\widetilde{{\bm{\theta}}}$ and $\widetilde{{\bm{\varphi }}}$, \eqref{L tilde} and \eqref{L-diag}, is diagonal, the correlation functions are standard. Incorporating the boundary conditions, \eqref{bc tilde}, results \cite{IVY:2013,*YY2014} in the $\widetilde{\theta}$-$\widetilde{\theta}$ and $\widetilde{\varphi} $-$\widetilde{\varphi} $ correlations of \eqref{CFtilde}, and the following antisymmetric correlations of $\widetilde{{\bm{\theta}}}$ and $\widetilde{{\bm{\varphi }}}$:
\begin{align}
-\langle\Delta{\widetilde{\bm\varphi}}(t)\otimes 2{\widetilde{\bm\theta}}^{\rm T}\!(t')\rangle &=
\langle{2\widetilde{\bm\theta}}(t)\otimes\Delta{\widetilde{\bm\varphi}}^{\rm T}\!(t')\rangle
%\notag\\&
=  ({\widetilde{\mathsf R}}-{\widetilde{\mathsf T}} )\ell%\,\ln(t-t')\,.
\end{align}
 with $\ell \equiv \ln(t-t') $. This results after rotation \eqref{M} in the correlations   of the original fields $ {{\bm{\theta}}}$ and $ {{\bm{\varphi }}}$ given in \eqref{CF} and their cross-correlation given below:
 \begin{align}\label{A2}
    \langle  {\bm\theta}(t)\otimes\Delta{\bm\varphi}^{\rm T}\!({t'})\rangle
&=  \phantom{-}\frac{ \Xi{\mathsf K}}{1+\Xi{\mathsf K}}\ell \,  , \\   \langle  {\Delta{\bm\varphi}\otimes\bm\theta}(t)^{\rm T}\!({t'})\rangle
&= - \frac{ {\mathsf K}\Xi }{1+{\mathsf K}\Xi}\ell\,.
 \end{align}
 {The above structure guarantees that the cross-correlations will not affect   correlation functions of linear combinations of the type ${\bm{a\cdot\theta}}+{\bm{b\cdot }}\Delta{\bm{\varphi }}$, and thus will not enter the RG dimensions calculated below.}

 In the physical limit described after \eqref{bc}  the boundary conditions for ${\bm{\theta}}$ are relevant in continuous channels and for ${\bm{\varphi }}$ in split channels.
 To take the limit, we relabel the channels so that  the first $n$ are continuous and the rest $N-n$ are split. In such a basis the Luttinger matrix and its inverse can be written as
\begin{align}\label{K-K}
    \begin{aligned}
{\mathsf K}&=\left(
              \begin{array}{cc}
                {\mathsf K}_{{\mathrm{cc}}} & {\mathsf K}_{{\mathrm{ci}}}  \\
                {\mathsf K}_{{\mathrm{ic}}}  & {\mathsf K}_{{\mathrm{ii}}} \\
              \end{array}
            \right)\,,&{\mathsf{K}}^{-1} =\begin{pmatrix}
                                                                         \overline{{\mathsf{K}}}_{\mathrm{cc}} &  \overline{{\mathsf{K}}}_{\mathrm{ci}} \\
                                                                          \overline{{\mathsf{K}}}_{\mathrm{ic}} &  \overline{{\mathsf{K}}}_{\mathrm{ii}} \\
                                                                       \end{pmatrix}
\end{aligned}
\end{align}
while  $
\Xi  \equiv\operatorname{diag}(\hat\xi_{\mathrm{c}}, \hat\xi_{\mathrm{i}})$ where in the physical limit all the elements of the diagonal $n\times n$ matrix $\hat\xi _{\mathrm{c}}$ go to zero, and all the elements of the diagonal $(N-n)\times ({N-n})$ matrix $\hat{\xi }_{\mathrm{i}}$ to infinity. Obviously, $ \overline{{\mathsf{K}}}_{\mathrm{cc}}\ne  {{\mathsf{K}}}_{\mathrm{cc}}^{-1} $, as the elements of the former matrix depend on all the elements of matrix ${\mathsf{K}}$. In these notations one finds that
\begin{align}\label{lim1}
    \begin{aligned}
\lim_{\xi}\left[{\mathsf K}^{-1}+\Xi\right]^{-1}&= \left(
                                                   \begin{array}{cc}
                                                    \overline{{\mathsf{K}}}_{\mathrm{cc}}   ^{\,\,-1} & 0 \\
                                                     0 & 0 \\
                                                   \end{array}
                                                 \right)\,,\\
\lim_{\xi}\left[{\mathsf K}+\Xi^{-1}\right]^{-1}&= \left(
                                                   \begin{array}{cc}
                                                     0 & 0 \\
                                                     0 &   {{\mathsf{K}}}_{\mathrm{ii}}  ^{\,-1} \\
                                                   \end{array}
                                                 \right)\,.
\end{aligned}
\end{align}
Thus in terms of the relabeled channels the right-hand sides of \eqref{CFa} and \eqref{CFb} go over to $ \overline{{\mathsf{K}}}_{\mathrm{cc}} ^{\,\,-1}\ell $ and $  {{\mathsf{K}}}_{\mathrm{ii}}  ^{\,-1}\ell $, respectively.

Using the relabeled channels, we rewrite the Lagrangian density of \eqref{pert}   as
\begin{align}\label{pert1}
{\cal L}_{\rm sc}&=\sum_{{\bm n}}v_{\bm n}\,e^{i{\bm n} \cdot{\bm\Phi}}+\mathrm{ c.c.}\,,& {\bm\Phi}&=
                                                                              \begin{pmatrix}
                                                                                2{\bm\theta} \\
\Delta{\bm\varphi} \\
                                                                              \end{pmatrix},
& {\bm n}&= \begin{pmatrix}
                            {\bm n}_c \\
                            {\bm n}_i \\
                        \end{pmatrix}\,,
\end{align}
where ${\bm n}_c$ and ${\bm n}_i$ are integer-valued vectors  belonging to the ${\mathrm{c}}$- and ${\mathrm{i}}$-subspaces, respectively, that describe the multiplicity of backscattering in the former and of  tunneling  in the latter. The correlation function of  fields ${\bm{\Phi }}$ is not contributed by the the off-diagonal correlation of \eqref{A2} and is obtained from \eqref{CF} in the limit \eqref{lim1} as follows:
\begin{equation}
\tfrac{1}{2}\,\langle {\bm\Phi}(t)\otimes{\bm\Phi}^{\rm T}\!({t'})\rangle=\begin{pmatrix}
                                                 \overline{{\mathsf{K}}}_{\mathrm{cc}}   ^{\,\,-1} & 0 \\
                                                  0 & {\mathsf K}_{{\mathrm{ii}}}^{-1} \\
                                               \end{pmatrix}=\begin{pmatrix}
                                                  \overline{{\mathsf{K}}}_{\mathrm{cc}}^{\phantom{-1}}\!\!\!\!     & 0 \\
                                                  0 & {\mathsf K}_{{\mathrm{ii}}}^{\phantom{-1}}\!\!\!\! \\
                                               \end{pmatrix}^{-1}\!\!\!\!.
\end{equation}
Therefore, the scaling dimension of each term in Lagrangian \eqref{pert1} can be written as
\begin{equation}\label{dim}
{\mathrm {dim}}\left[v_{\bm n}\,e^{i{\bm n}^{\rm T}{\bm\Phi}}\right]={\bm n}^{\rm T}\left(
                                                \begin{array}{cc}
                                                  {\bar{\mathsf K}}_{\rm cc} & 0 \\
                                                  0 & {\mathsf K}_{\rm ii} \\
                                                \end{array}
                                              \right)^{-1}{\bm n}\,.
\end{equation}
Now we use the projector operators of \eqref{P} to restore the original  numbering of the channels which gives
\begin{align}
{ \overline{{\mathsf K}}}_{\rm cc}&\to   {\mathsf P}_{\rm c}\,{\mathsf K}^{-1}\,{\mathsf P}_{\rm c}\,,&
{\mathsf K}_{\rm ii}&\to  {\mathsf P}_{\rm i}\,{\mathsf K}\,{\mathsf P}_{\rm i}\,,
\end{align}
Combining this with \eqref{dim} results in \eqref{Delta} in the main text.

\section{the shortest vector problem}
Finding the minimum of a quadratic form built on integer-valued vectors is equivalent to finding the shortest vector connecting nodes on a lattice. Although this problem in its completeness is known to be computationally hard  \cite{SVP:1,*SVP} determining the sufficient condition for the shortest vector to be not an elementary lattice vector is straightforward. This is all we need to define the parametric region in which one-particle scattering is not necessarily RG-dominant.

The elements of the $2\times 2$ Luttinger matrix ${\mathsf K}$ in the Gram representation are written as   $\{{K_{ij}}\}=\bm g_i\cdot \bm g_j $, where $|{\bm{g}}_i|=\sqrt{K_{ii} }$, while the angle $\gamma=\widehat{{\bm{g}}_1{\bm{g}}_2}$ is given by \eqref{gamma}.
Then one has to find the minimum of $|\bm{G}|^2$ where ${\bm{G}}=n_1{\bm{g}}_1+n_2{\bm{g}}_2 $, i.e.\ the minimal distance between two nodes on a two-dimensional lattice spanned by the   basis vectors ${\bm{g}}_{1,2} $. For a rectangular lattice ({$\cos \gamma=0$}) the solution is the shortest lattice spacing, corresponding to $n_1=0\,,\,n_2=\pm 1$ (assuming $g_1>g_2$).

 \begin{figure}\vspace{12pt}

 \centerline{
 {\includegraphics[width=.9\columnwidth]{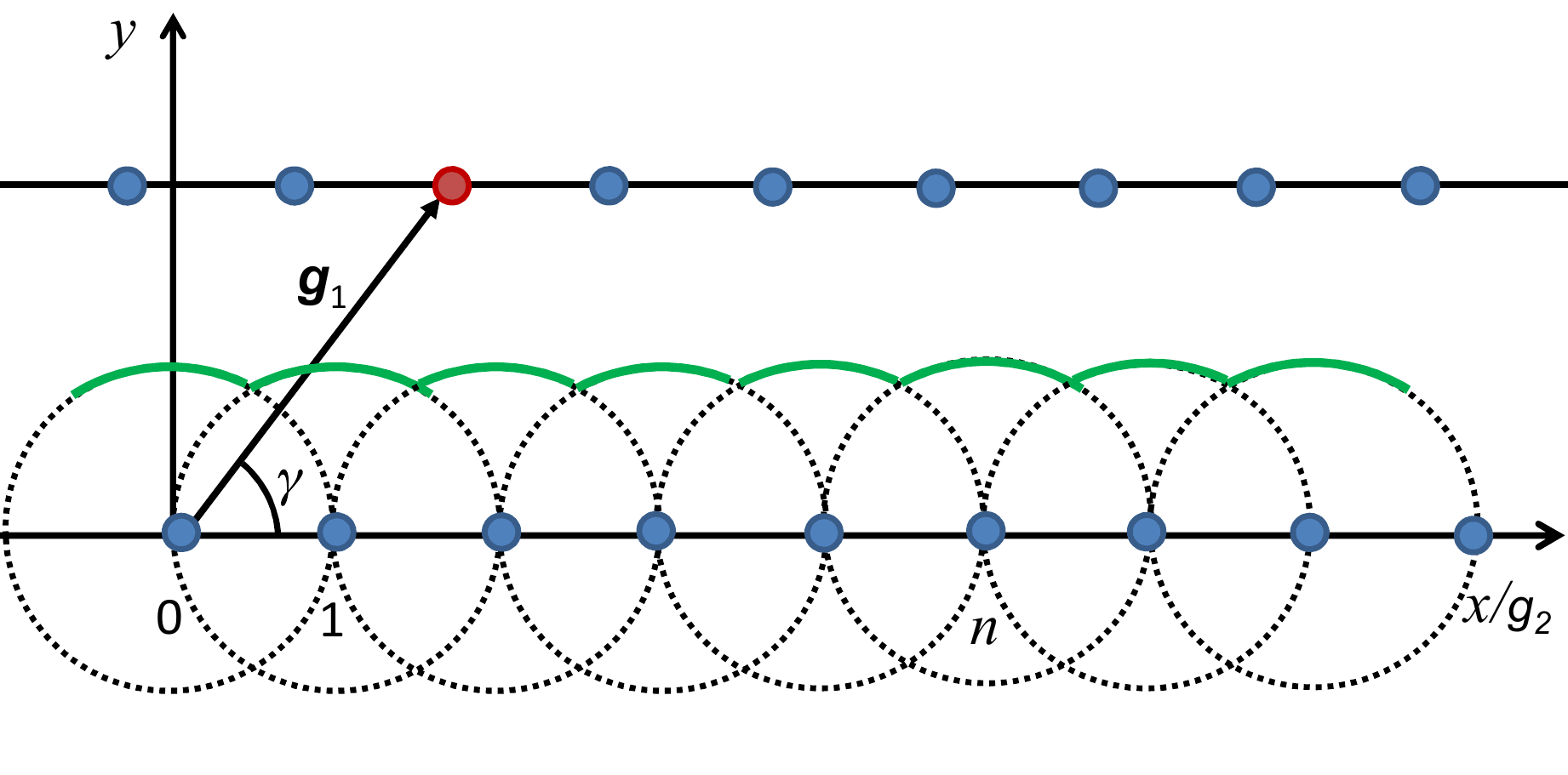}}
  }
\caption{The SVP illustration. With $\gamma$ decreasing, the nearest horizontal chains of the lattice become closer. At some critical angle, the upper chain crosses the boundary made by circles of radius $g_2$ around each node of the lower chain. Then the shortest distance between the nodes of the upper and lower chains is less than the length of the shortest basis vector $\bm g_2$.}
\label{4}
\end{figure}

On   decreasing the lattice angle $\gamma$ with $g_1 \geqslant g_2$ being constant,  the horizontal lattice chains become closer as illustrated in Fig.~\ref4. We draw there the circles of radius $g_2$ centered at the lattice nodes on the low horizontal chain (with $y=0$). Measuring all lengths in units of $g_2$, the $y$ coordinate of the upper boundary of these circles can be written as $y=\sqrt{1-\{{R\cos \gamma}\}^2 }$, where $R\equiv g_1/g_2 \geqslant 1$ and $\{{A}\} $ is the distance of $A$ to the closest integer $n$ (so that $-\frac{1}{2}\leqslant\{{A}\} \leqslant\frac{1}{2} $). When the end of basis vector $\bm g_1$  touches this boundary, the distance between the zeroth node
 of the upper and the $n^{{\mathrm{th}}} $ node of the lower chains  equals  $g_2$  and becomes smaller with $\gamma$ further decreasing -- this is where the $n$-particle scattering becomes more RG-relevant than the one-particle. As the $x$ coordinate of $\bm g_1$  equals $R\sin \gamma$, the condition for this not to happen for any $n$ is
 \begin{equation}
R^2\sin^2\gamma +\{R\cos\gamma\}^2 > 1\,,
\end{equation}
Since $R\equiv g_1/g_2\geqslant1$ and $\{R\cos\gamma\}^2\leqslant\frac{1}{4}$,  the inequality  is satisfied for any  $R$ when $\cos \gamma<\frac{1}{2}$. {When the inequality fails, the multiplicity $n$ of the scattering process which is more RG relevant than one-particle scattering is given by $n=\left[R\cos \gamma\right]+1$ where $\left[A\right]$ is an integer closest to $A$. Thus, depending on the ratio $R$, it could arbitrary large. For the important case of $R=1$ (considered in Section \ref{topo}), it is the physically relevant  \cite{KLY:16d} two-particle scattering that becomes more RG relevant than one-particle for $\cos \gamma<\frac{1}{2}$.}

\end{appendix}

%\bibliography{my}\end{document}

\begin{thebibliography}{39}%
\makeatletter
\providecommand \@ifxundefined [1]{%
 \@ifx{#1\undefined}
}%
\providecommand \@ifnum [1]{%
 \ifnum #1\expandafter \@firstoftwo
 \else \expandafter \@secondoftwo
 \fi
}%
\providecommand \@ifx [1]{%
 \ifx #1\expandafter \@firstoftwo
 \else \expandafter \@secondoftwo
 \fi
}%
\providecommand \natexlab [1]{#1}%
\providecommand \enquote  [1]{``#1''}%
\providecommand \bibnamefont  [1]{#1}%
\providecommand \bibfnamefont [1]{#1}%
\providecommand \citenamefont [1]{#1}%
\providecommand \href@noop [0]{\@secondoftwo}%
\providecommand \href [0]{\begingroup \@sanitize@url \@href}%
\providecommand \@href[1]{\@@startlink{#1}\@@href}%
\providecommand \@@href[1]{\endgroup#1\@@endlink}%
\providecommand \@sanitize@url [0]{\catcode `\\12\catcode `\$12\catcode
  `\&12\catcode `\#12\catcode `\^12\catcode `\_12\catcode `\%12\relax}%
\providecommand \@@startlink[1]{}%
\providecommand \@@endlink[0]{}%
\providecommand \url  [0]{\begingroup\@sanitize@url \@url }%
\providecommand \@url [1]{\endgroup\@href {#1}{\urlprefix }}%
\providecommand \urlprefix  [0]{URL }%
\providecommand \Eprint [0]{\href }%
\providecommand \doibase [0]{http://dx.doi.org/}%
\providecommand \selectlanguage [0]{\@gobble}%
\providecommand \bibinfo  [0]{\@secondoftwo}%
\providecommand \bibfield  [0]{\@secondoftwo}%
\providecommand \translation [1]{[#1]}%
\providecommand \BibitemOpen [0]{}%
\providecommand \bibitemStop [0]{}%
\providecommand \bibitemNoStop [0]{.\EOS\space}%
\providecommand \EOS [0]{\spacefactor3000\relax}%
\providecommand \BibitemShut  [1]{\csname bibitem#1\endcsname}%
\let\auto@bib@innerbib\@empty
%</preamble>
\bibitem [{\citenamefont {Teo}\ and\ \citenamefont {Kane}(2014)}]{TK2014}%
  \BibitemOpen
  \bibfield  {author} {\bibinfo {author} {\bibfnamefont {Jeffrey C~Y}\
  \bibnamefont {Teo}}\ and\ \bibinfo {author} {\bibfnamefont {C~L}\
  \bibnamefont {Kane}},\ }\bibfield  {title} {\enquote {\bibinfo {title} {From
  {{Luttinger}} liquid to non-{Abelian} quantum {Hall} states},}\ }\href@noop
  {} {\bibfield  {journal} {\bibinfo  {journal} {Phys. Rev. B}\ }\textbf
  {\bibinfo {volume} {89}},\ \bibinfo {pages} {085101} (\bibinfo {year}
  {2014})}\BibitemShut {NoStop}%
\bibitem [{\citenamefont {Sondhi}\ and\ \citenamefont
  {Yang}(2001)}]{Sondhi2001}%
  \BibitemOpen
  \bibfield  {author} {\bibinfo {author} {\bibfnamefont {S.~L.}\ \bibnamefont
  {Sondhi}}\ and\ \bibinfo {author} {\bibfnamefont {Kun}\ \bibnamefont
  {Yang}},\ }\bibfield  {title} {\enquote {\bibinfo {title} {Sliding phases via
  magnetic fields},}\ }\href {\doibase 10.1103/PhysRevB.63.054430} {\bibfield
  {journal} {\bibinfo  {journal} {Phys. Rev. B}\ }\textbf {\bibinfo {volume}
  {63}},\ \bibinfo {pages} {054430} (\bibinfo {year} {2001})}\BibitemShut
  {NoStop}%
\bibitem [{\citenamefont {Kane}\ \emph {et~al.}({2002})\citenamefont {Kane},
  \citenamefont {Mukhopadhyay},\ and\ \citenamefont {Lubensky}}]{Kane2002}%
  \BibitemOpen
  \bibfield  {author} {\bibinfo {author} {\bibfnamefont {C~L}\ \bibnamefont
  {Kane}}, \bibinfo {author} {\bibfnamefont {R}~\bibnamefont {Mukhopadhyay}}, \
  and\ \bibinfo {author} {\bibfnamefont {T~C}\ \bibnamefont {Lubensky}},\
  }\bibfield  {title} {\enquote {\bibinfo {title} {Fractional quantum {Hall}
  effect in an array of quantum wires},}\ }\href@noop {} {\bibfield  {journal}
  {\bibinfo  {journal} {{Phys. Rev. Lett.}}\ }\textbf {\bibinfo {volume}
  {{88}}},\ \bibinfo {pages} {{036401}} (\bibinfo {year} {{2002}})}\BibitemShut
  {NoStop}%
\bibitem [{\citenamefont {O'Hern}\ \emph {et~al.}(1999)\citenamefont {O'Hern},
  \citenamefont {Lubensky},\ and\ \citenamefont {Toner}}]{OHern1999}%
  \BibitemOpen
  \bibfield  {author} {\bibinfo {author} {\bibfnamefont {C.~S.}\ \bibnamefont
  {O'Hern}}, \bibinfo {author} {\bibfnamefont {T.~C.}\ \bibnamefont
  {Lubensky}}, \ and\ \bibinfo {author} {\bibfnamefont {J.}~\bibnamefont
  {Toner}},\ }\bibfield  {title} {\enquote {\bibinfo {title} {Sliding phases in
  $\mathit{XY}$ models, crystals, and cationic lipid-{DNA} complexes},}\ }\href
  {\doibase 10.1103/PhysRevLett.83.2745} {\bibfield  {journal} {\bibinfo
  {journal} {Phys. Rev. Lett.}\ }\textbf {\bibinfo {volume} {83}},\ \bibinfo
  {pages} {2745--2748} (\bibinfo {year} {1999})}\BibitemShut {NoStop}%
\bibitem [{\citenamefont {Vishwanath}\ and\ \citenamefont
  {Carpentier}(2001)}]{Vishwanath2001}%
  \BibitemOpen
  \bibfield  {author} {\bibinfo {author} {\bibfnamefont {Ashvin}\ \bibnamefont
  {Vishwanath}}\ and\ \bibinfo {author} {\bibfnamefont {David}\ \bibnamefont
  {Carpentier}},\ }\bibfield  {title} {\enquote {\bibinfo {title}
  {Two-dimensional anisotropic non-{Fermi}-liquid phase of coupled {Luttinger}
  liquids},}\ }\href {\doibase 10.1103/PhysRevLett.86.676} {\bibfield
  {journal} {\bibinfo  {journal} {Phys. Rev. Lett.}\ }\textbf {\bibinfo
  {volume} {86}},\ \bibinfo {pages} {676--679} (\bibinfo {year}
  {2001})}\BibitemShut {NoStop}%
\bibitem [{\citenamefont {Mukhopadhyay}\ \emph {et~al.}(2001)\citenamefont
  {Mukhopadhyay}, \citenamefont {Kane},\ and\ \citenamefont
  {Lubensky}}]{MKL2001}%
  \BibitemOpen
  \bibfield  {author} {\bibinfo {author} {\bibfnamefont {Ranjan}\ \bibnamefont
  {Mukhopadhyay}}, \bibinfo {author} {\bibfnamefont {C.~L.}\ \bibnamefont
  {Kane}}, \ and\ \bibinfo {author} {\bibfnamefont {T.~C.}\ \bibnamefont
  {Lubensky}},\ }\bibfield  {title} {\enquote {\bibinfo {title} {Crossed
  sliding {Luttinger} liquid phase},}\ }\href {\doibase
  10.1103/PhysRevB.63.081103} {\bibfield  {journal} {\bibinfo  {journal} {Phys.
  Rev. B}\ }\textbf {\bibinfo {volume} {63}},\ \bibinfo {pages} {081103}
  (\bibinfo {year} {2001})}\BibitemShut {NoStop}%
\bibitem [{\citenamefont {Cazalilla}\ and\ \citenamefont {Ho}(2003)}]{HoCaz}%
  \BibitemOpen
  \bibfield  {author} {\bibinfo {author} {\bibfnamefont {M.~A.}\ \bibnamefont
  {Cazalilla}}\ and\ \bibinfo {author} {\bibfnamefont {A.~F.}\ \bibnamefont
  {Ho}},\ }\bibfield  {title} {\enquote {\bibinfo {title} {Instabilities in
  binary mixtures of one-dimensional quantum degenerate gases},}\ }\href
  {\doibase 10.1103/PhysRevLett.91.150403} {\bibfield  {journal} {\bibinfo
  {journal} {Phys. Rev. Lett.}\ }\textbf {\bibinfo {volume} {91}},\ \bibinfo
  {pages} {150403} (\bibinfo {year} {2003})}\BibitemShut {NoStop}%
\bibitem [{\citenamefont {Mathey}\ \emph {et~al.}(2004)\citenamefont {Mathey},
  \citenamefont {Wang}, \citenamefont {Hofstetter}, \citenamefont {Lukin},\
  and\ \citenamefont {Demler}}]{FB-Mat&&LukDem}%
  \BibitemOpen
  \bibfield  {author} {\bibinfo {author} {\bibfnamefont {L.}~\bibnamefont
  {Mathey}}, \bibinfo {author} {\bibfnamefont {D.-W.}\ \bibnamefont {Wang}},
  \bibinfo {author} {\bibfnamefont {W.}~\bibnamefont {Hofstetter}}, \bibinfo
  {author} {\bibfnamefont {M.~D.}\ \bibnamefont {Lukin}}, \ and\ \bibinfo
  {author} {\bibfnamefont {Eugene}\ \bibnamefont {Demler}},\ }\bibfield
  {title} {\enquote {\bibinfo {title} {{Luttinger} liquid of polarons in
  one-dimensional boson-fermion mixtures},}\ }\href@noop {} {\bibfield
  {journal} {\bibinfo  {journal} {Phys. Rev. Lett.}\ }\textbf {\bibinfo
  {volume} {93}},\ \bibinfo {pages} {120404} (\bibinfo {year}
  {2004})}\BibitemShut {NoStop}%
\bibitem [{\citenamefont {Cr\'epin}\ \emph {et~al.}(2010)\citenamefont
  {Cr\'epin}, \citenamefont {Zar\'and},\ and\ \citenamefont
  {Simon}}]{BF-Simon:10}%
  \BibitemOpen
  \bibfield  {author} {\bibinfo {author} {\bibfnamefont {F}~\bibnamefont
  {Cr\'epin}}, \bibinfo {author} {\bibfnamefont {Gergely}\ \bibnamefont
  {Zar\'and}}, \ and\ \bibinfo {author} {\bibfnamefont {Pascal}\ \bibnamefont
  {Simon}},\ }\bibfield  {title} {\enquote {\bibinfo {title} {Disordered
  one-dimensional {Bose}-{Fermi} mixtures: The {Bose}-{Fermi} glass},}\ }\href
  {\doibase 10.1103/PhysRevLett.105.115301} {\bibfield  {journal} {\bibinfo
  {journal} {Phys. Rev. Lett.}\ }\textbf {\bibinfo {volume} {105}},\ \bibinfo
  {pages} {115301} (\bibinfo {year} {2010})}\BibitemShut {NoStop}%
\bibitem [{\citenamefont {Cr\'epin}\ \emph {et~al.}(2012)\citenamefont
  {Cr\'epin}, \citenamefont {Zar\'and},\ and\ \citenamefont
  {Simon}}]{BF-Simon:12}%
  \BibitemOpen
  \bibfield  {author} {\bibinfo {author} {\bibfnamefont {F}~\bibnamefont
  {Cr\'epin}}, \bibinfo {author} {\bibfnamefont {Gergely}\ \bibnamefont
  {Zar\'and}}, \ and\ \bibinfo {author} {\bibfnamefont {Pascal}\ \bibnamefont
  {Simon}},\ }\bibfield  {title} {\enquote {\bibinfo {title} {Mixtures of
  ultracold atoms in one-dimensional disordered potentials},}\ }\href {\doibase
  10.1103/PhysRevA.85.023625} {\bibfield  {journal} {\bibinfo  {journal} {Phys.
  Rev. A}\ }\textbf {\bibinfo {volume} {85}},\ \bibinfo {pages} {023625}
  (\bibinfo {year} {2012})}\BibitemShut {NoStop}%
\bibitem [{\citenamefont {San-Jose}\ \emph {et~al.}(2005)\citenamefont
  {San-Jose}, \citenamefont {Guinea},\ and\ \citenamefont {Martin}}]{martin05}%
  \BibitemOpen
  \bibfield  {author} {\bibinfo {author} {\bibfnamefont {Pablo}\ \bibnamefont
  {San-Jose}}, \bibinfo {author} {\bibfnamefont {Francisco}\ \bibnamefont
  {Guinea}}, \ and\ \bibinfo {author} {\bibfnamefont {Thierry}\ \bibnamefont
  {Martin}},\ }\bibfield  {title} {\enquote {\bibinfo {title} {Electron
  backscattering from dynamical impurities in a {Luttinger} liquid},}\ }\href
  {\doibase 10.1103/PhysRevB.72.165427} {\bibfield  {journal} {\bibinfo
  {journal} {Phys. Rev. B}\ }\textbf {\bibinfo {volume} {72}},\ \bibinfo
  {pages} {165427} (\bibinfo {year} {2005})}\BibitemShut {NoStop}%
\bibitem [{\citenamefont {Galda}\ \emph
  {et~al.}(2011{\natexlab{a}})\citenamefont {Galda}, \citenamefont
  {Yurkevich},\ and\ \citenamefont {Lerner}}]{GYL:2011}%
  \BibitemOpen
  \bibfield  {author} {\bibinfo {author} {\bibfnamefont {Alexey}\ \bibnamefont
  {Galda}}, \bibinfo {author} {\bibfnamefont {Igor~V.}\ \bibnamefont
  {Yurkevich}}, \ and\ \bibinfo {author} {\bibfnamefont {Igor~V.}\ \bibnamefont
  {Lerner}},\ }\bibfield  {title} {\enquote {\bibinfo {title} {Impurity
  scattering in a {Luttinger} liquid with electron-phonon coupling},}\ }\href
  {\doibase 10.1103/PhysRevB.83.041106} {\bibfield  {journal} {\bibinfo
  {journal} {Phys. Rev. B}\ }\textbf {\bibinfo {volume} {83}},\ \bibinfo
  {pages} {R041106} (\bibinfo {year} {2011}{\natexlab{a}})}\BibitemShut
  {NoStop}%
\bibitem [{\citenamefont {Galda}\ \emph
  {et~al.}(2011{\natexlab{b}})\citenamefont {Galda}, \citenamefont
  {Yurkevich},\ and\ \citenamefont {Lerner}}]{GYL:2011a}%
  \BibitemOpen
  \bibfield  {author} {\bibinfo {author} {\bibfnamefont {A.}~\bibnamefont
  {Galda}}, \bibinfo {author} {\bibfnamefont {I.~V.}\ \bibnamefont
  {Yurkevich}}, \ and\ \bibinfo {author} {\bibfnamefont {I.~V.}\ \bibnamefont
  {Lerner}},\ }\bibfield  {title} {\enquote {\bibinfo {title} {Effect of
  electron-phonon coupling on transmission through {Luttinger} liquid
  hybridized with resonant level},}\ }\href {\doibase
  10.1209/0295-5075/93/17009} {\bibfield  {journal} {\bibinfo  {journal} {EPL}\
  }\textbf {\bibinfo {volume} {93}},\ \bibinfo {pages} {17009} (\bibinfo {year}
  {2011}{\natexlab{b}})}\BibitemShut {NoStop}%
\bibitem [{\citenamefont {Yurkevich}\ \emph {et~al.}(2013)\citenamefont
  {Yurkevich}, \citenamefont {Galda}, \citenamefont {Yevtushenko},\ and\
  \citenamefont {Lerner}}]{YGYL:2013}%
  \BibitemOpen
  \bibfield  {author} {\bibinfo {author} {\bibfnamefont {Igor~V.}\ \bibnamefont
  {Yurkevich}}, \bibinfo {author} {\bibfnamefont {Alexey}\ \bibnamefont
  {Galda}}, \bibinfo {author} {\bibfnamefont {Oleg~M.}\ \bibnamefont
  {Yevtushenko}}, \ and\ \bibinfo {author} {\bibfnamefont {Igor~V.}\
  \bibnamefont {Lerner}},\ }\bibfield  {title} {\enquote {\bibinfo {title}
  {Duality of weak and strong scatterer in a {Luttinger} liquid coupled to
  massless bosons},}\ }\href {\doibase 10.1103/PhysRevLett.110.136405}
  {\bibfield  {journal} {\bibinfo  {journal} {Phys. Rev. Lett.}\ }\textbf
  {\bibinfo {volume} {110}},\ \bibinfo {pages} {136405} (\bibinfo {year}
  {2013})}\BibitemShut {NoStop}%
\bibitem [{\citenamefont {Yurkevich}(2013)}]{IVY:2013}%
  \BibitemOpen
  \bibfield  {author} {\bibinfo {author} {\bibfnamefont {Igor~V.}\ \bibnamefont
  {Yurkevich}},\ }\bibfield  {title} {\enquote {\bibinfo {title} {Duality in
  multi-channel {Luttinger} liquid with local scatterer},}\ }\href@noop {}
  {\bibfield  {journal} {\bibinfo  {journal} {EPL}\ }\textbf {\bibinfo {volume}
  {104}},\ \bibinfo {pages} {37004} (\bibinfo {year} {2013})}\BibitemShut
  {NoStop}%
\bibitem [{\citenamefont {Yurkevich}\ and\ \citenamefont
  {Yevtushenko}(2014)}]{YY2014}%
  \BibitemOpen
  \bibfield  {author} {\bibinfo {author} {\bibfnamefont {Igor~V.}\ \bibnamefont
  {Yurkevich}}\ and\ \bibinfo {author} {\bibfnamefont {Oleg~M.}\ \bibnamefont
  {Yevtushenko}},\ }\bibfield  {title} {\enquote {\bibinfo {title} {Universal
  duality in a {Luttinger} liquid coupled to a generic environment},}\ }\href
  {\doibase 10.1103/PhysRevB.90.115411} {\bibfield  {journal} {\bibinfo
  {journal} {Phys. Rev. B}\ }\textbf {\bibinfo {volume} {90}},\ \bibinfo
  {pages} {115411} (\bibinfo {year} {2014})}\BibitemShut {NoStop}%
\bibitem [{\citenamefont {Hou}\ \emph {et~al.}(2009)\citenamefont {Hou},
  \citenamefont {Kim},\ and\ \citenamefont {Chamon}}]{PRL09-Chamon}%
  \BibitemOpen
  \bibfield  {author} {\bibinfo {author} {\bibfnamefont {Chang-Yu}\
  \bibnamefont {Hou}}, \bibinfo {author} {\bibfnamefont {Eun-Ah}\ \bibnamefont
  {Kim}}, \ and\ \bibinfo {author} {\bibfnamefont {Claudio}\ \bibnamefont
  {Chamon}},\ }\bibfield  {title} {\enquote {\bibinfo {title} {Corner junction
  as a probe of helical edge states},}\ }\href {\doibase
  10.1103/PhysRevLett.102.076602} {\bibfield  {journal} {\bibinfo  {journal}
  {Phys. Rev. Lett.}\ }\textbf {\bibinfo {volume} {102}},\ \bibinfo {pages}
  {076602} (\bibinfo {year} {2009})}\BibitemShut {NoStop}%
\bibitem [{\citenamefont {Santos}\ and\ \citenamefont
  {Gutman}(2015)}]{Santos2015}%
  \BibitemOpen
  \bibfield  {author} {\bibinfo {author} {\bibfnamefont {Raul~A.}\ \bibnamefont
  {Santos}}\ and\ \bibinfo {author} {\bibfnamefont {D.~B.}\ \bibnamefont
  {Gutman}},\ }\bibfield  {title} {\enquote {\bibinfo {title}
  {Interaction-protected topological insulators with time reversal symmetry},}\
  }\href {\doibase 10.1103/PhysRevB.92.075135} {\bibfield  {journal} {\bibinfo
  {journal} {Phys. Rev. B}\ }\textbf {\bibinfo {volume} {92}},\ \bibinfo
  {pages} {075135} (\bibinfo {year} {2015})}\BibitemShut {NoStop}%
\bibitem [{\citenamefont {Kane}\ and\ \citenamefont
  {Fisher}(1992{\natexlab{a}})}]{KaneFis:92a}%
  \BibitemOpen
  \bibfield  {author} {\bibinfo {author} {\bibfnamefont {C~L}\ \bibnamefont
  {Kane}}\ and\ \bibinfo {author} {\bibfnamefont {M~P~A}\ \bibnamefont
  {Fisher}},\ }\bibfield  {title} {\enquote {\bibinfo {title} {Transport in a
  one-channel {Luttinger} liquid},}\ }\href@noop {} {\bibfield  {journal}
  {\bibinfo  {journal} {Phys. Rev. Lett.}\ }\textbf {\bibinfo {volume} {68}},\
  \bibinfo {pages} {1220} (\bibinfo {year} {1992}{\natexlab{a}})}\BibitemShut
  {NoStop}%
\bibitem [{\citenamefont {Kane}\ and\ \citenamefont
  {Fisher}(1992{\natexlab{b}})}]{KaneFis:92b}%
  \BibitemOpen
  \bibfield  {author} {\bibinfo {author} {\bibfnamefont {C~L}\ \bibnamefont
  {Kane}}\ and\ \bibinfo {author} {\bibfnamefont {M~P~A}\ \bibnamefont
  {Fisher}},\ }\bibfield  {title} {\enquote {\bibinfo {title} {Resonant
  tunneling in an interacting one-dimensional electron gas},}\ }\href@noop {}
  {\bibfield  {journal} {\bibinfo  {journal} {Phys. Rev. {\rm B}}\ }\textbf
  {\bibinfo {volume} {46}},\ \bibinfo {pages} {7268(R)} (\bibinfo {year}
  {1992}{\natexlab{b}})}\BibitemShut {NoStop}%
\bibitem [{\citenamefont {Kane}\ and\ \citenamefont
  {Fisher}(1992{\natexlab{c}})}]{KF:92b}%
  \BibitemOpen
  \bibfield  {author} {\bibinfo {author} {\bibfnamefont {C.~L.}\ \bibnamefont
  {Kane}}\ and\ \bibinfo {author} {\bibfnamefont {Matthew P.~A.}\ \bibnamefont
  {Fisher}},\ }\bibfield  {title} {\enquote {\bibinfo {title} {Transmission
  through barriers and resonant tunneling in an interacting one-dimensional
  electron gas},}\ }\href {\doibase 10.1103/PhysRevB.46.15233} {\bibfield
  {journal} {\bibinfo  {journal} {Phys. Rev. {\rm B}}\ }\textbf {\bibinfo
  {volume} {46}},\ \bibinfo {pages} {15233--15262} (\bibinfo {year}
  {1992}{\natexlab{c}})}\BibitemShut {NoStop}%
\bibitem [{\citenamefont {Maslov}\ and\ \citenamefont
  {Stone}(1995)}]{MasStone:95}%
  \BibitemOpen
  \bibfield  {author} {\bibinfo {author} {\bibfnamefont {D~L}\ \bibnamefont
  {Maslov}}\ and\ \bibinfo {author} {\bibfnamefont {M}~\bibnamefont {Stone}},\
  }\bibfield  {title} {\enquote {\bibinfo {title} {Landauer conductance of
  {Luttinger} liquids with leads},}\ }\href@noop {} {\bibfield  {journal}
  {\bibinfo  {journal} {Phys. Rev. {\rm B}}\ }\textbf {\bibinfo {volume}
  {52}},\ \bibinfo {pages} {R5539} (\bibinfo {year} {1995})}\BibitemShut
  {NoStop}%
\bibitem [{\citenamefont {Ponomarenko}(1995)}]{Ponomarenko:95}%
  \BibitemOpen
  \bibfield  {author} {\bibinfo {author} {\bibfnamefont {V.~V.}\ \bibnamefont
  {Ponomarenko}},\ }\bibfield  {title} {\enquote {\bibinfo {title}
  {Renormalization of the one-dimensional conductance in the {Luttinger}-liquid
  model},}\ }\href@noop {} {\bibfield  {journal} {\bibinfo  {journal} {Phys.
  Rev. {\rm B}}\ }\textbf {\bibinfo {volume} {52}},\ \bibinfo {pages}
  {R8666--R8667} (\bibinfo {year} {1995})}\BibitemShut {NoStop}%
\bibitem [{\citenamefont {Safi}\ and\ \citenamefont
  {Schulz}(1995)}]{SafiSch:95}%
  \BibitemOpen
  \bibfield  {author} {\bibinfo {author} {\bibfnamefont {I.}~\bibnamefont
  {Safi}}\ and\ \bibinfo {author} {\bibfnamefont {H.~J.}\ \bibnamefont
  {Schulz}},\ }\bibfield  {title} {\enquote {\bibinfo {title} {Transport in an
  inhomogeneous interacting one-dimensional system},}\ }\href@noop {}
  {\bibfield  {journal} {\bibinfo  {journal} {Phys. Rev. {\rm B}}\ }\textbf
  {\bibinfo {volume} {52}},\ \bibinfo {pages} {R17040--R17043} (\bibinfo {year}
  {1995})}\BibitemShut {NoStop}%
\bibitem [{\citenamefont {Giamarchi}\ and\ \citenamefont
  {Schulz}(1988)}]{GS1988}%
  \BibitemOpen
  \bibfield  {author} {\bibinfo {author} {\bibfnamefont {T.}~\bibnamefont
  {Giamarchi}}\ and\ \bibinfo {author} {\bibfnamefont {H.~J.}\ \bibnamefont
  {Schulz}},\ }\bibfield  {title} {\enquote {\bibinfo {title} {Anderson
  localization and interactions in one-dimensional metals},}\ }\href {\doibase
  10.1103/PhysRevB.37.325} {\bibfield  {journal} {\bibinfo  {journal} {Phys.
  Rev. B}\ }\textbf {\bibinfo {volume} {37}},\ \bibinfo {pages} {325--340}
  (\bibinfo {year} {1988})}\BibitemShut {NoStop}%
\bibitem [{\citenamefont {Basko}\ \emph {et~al.}(2006)\citenamefont {Basko},
  \citenamefont {Aleiner},\ and\ \citenamefont {Altshuler}}]{BAA1}%
  \BibitemOpen
  \bibfield  {author} {\bibinfo {author} {\bibfnamefont {D~M}\ \bibnamefont
  {Basko}}, \bibinfo {author} {\bibfnamefont {I~L}\ \bibnamefont {Aleiner}}, \
  and\ \bibinfo {author} {\bibfnamefont {B~L}\ \bibnamefont {Altshuler}},\
  }\bibfield  {title} {\enquote {\bibinfo {title} {Metal-insulator transition
  in a weakly interacting many-electron system with localized single-particle
  states},}\ }\href@noop {} {\bibfield  {journal} {\bibinfo  {journal} {Ann.
  Phys.}\ }\textbf {\bibinfo {volume} {321}},\ \bibinfo {pages} {1126}
  (\bibinfo {year} {2006})}\BibitemShut {NoStop}%
\bibitem [{\citenamefont {Basko}\ \emph {et~al.}(2007)\citenamefont {Basko},
  \citenamefont {Aleiner},\ and\ \citenamefont {Altshuler}}]{BAA2}%
  \BibitemOpen
  \bibfield  {author} {\bibinfo {author} {\bibfnamefont {D.~M.}\ \bibnamefont
  {Basko}}, \bibinfo {author} {\bibfnamefont {I.~L.}\ \bibnamefont {Aleiner}},
  \ and\ \bibinfo {author} {\bibfnamefont {B.~L.}\ \bibnamefont {Altshuler}},\
  }\bibfield  {title} {\enquote {\bibinfo {title} {Possible experimental
  manifestations of the many-body localization},}\ }\href {\doibase
  10.1103/PhysRevB.76.052203} {\bibfield  {journal} {\bibinfo  {journal} {Phys.
  Rev. {\rm B}}\ }\textbf {\bibinfo {volume} {76}},\ \bibinfo {pages} {052203}
  (\bibinfo {year} {2007})}\BibitemShut {NoStop}%
\bibitem [{\citenamefont {Aleiner}\ \emph {et~al.}(2010)\citenamefont
  {Aleiner}, \citenamefont {Altshuler},\ and\ \citenamefont
  {Shlyapnikov}}]{AlAlS:2010}%
  \BibitemOpen
  \bibfield  {author} {\bibinfo {author} {\bibfnamefont {I.~L.}\ \bibnamefont
  {Aleiner}}, \bibinfo {author} {\bibfnamefont {B.~L.}\ \bibnamefont
  {Altshuler}}, \ and\ \bibinfo {author} {\bibfnamefont {G.~V.}\ \bibnamefont
  {Shlyapnikov}},\ }\bibfield  {title} {\enquote {\bibinfo {title} {{A
  finite-temperature phase transition for disordered weakly interacting bosons
  in one dimension}},}\ }\href {\doibase 10.1038/nphys1758} {\bibfield
  {journal} {\bibinfo  {journal} {Nature Phys.}\ }\textbf {\bibinfo {volume}
  {6}},\ \bibinfo {pages} {900} (\bibinfo {year} {2010})}\BibitemShut {NoStop}%
\bibitem [{KLY({\natexlab{a}})}]{KLY:16d}%
  \BibitemOpen
  \href@noop {} {} \bibinfo {note} {{The most RG-relevant
  scattering corresponds to the configuration with the lowest scaling
  dimension, see Eq.~\eqref{Delta}. However, bare amplitudes of multiparticle
  scattering are proportional to the appropriate power of a small parameter so
  that the regime where the most relevant multipatricle scattering dominates
  might only be reached at very lowe temperatures.},}\BibitemShut {NoStop}%
\bibitem [{\citenamefont {Haldane}(1981)}]{HALDANE:81}%
  \BibitemOpen
  \bibfield  {author} {\bibinfo {author} {\bibfnamefont {F~D~M}\ \bibnamefont
  {Haldane}},\ }\bibfield  {title} {\enquote {\bibinfo {title} {{Luttinger}
  liquid theory of one-dimensional quantum fluids},}\ }\href@noop {} {\bibfield
   {journal} {\bibinfo  {journal} {J. Phys. {\rm C}}\ }\textbf {\bibinfo
  {volume} {14}},\ \bibinfo {pages} {2585} (\bibinfo {year}
  {1981})}\BibitemShut {NoStop}%
\bibitem [{KLY({\natexlab{b}})}]{KLY:16a}%
  \BibitemOpen
  \href@noop {} {} \bibinfo {note} {{In the present case,
  both $\det {\mathsf{A}}\equiv \det{\mathsf{V}}_\theta $ and $\det
  {\mathsf{B}}\equiv \det{\mathsf{V}}_\varphi $ must be positive to avoid
  instabilities of the Wentzel--Bardeen type originally found
  \cite{Wentzel,*Bardeen:51} for the electron-phonon interaction in $1D$.
  }}\BibitemShut {NoStop}%
\bibitem [{KLY({\natexlab{c}})}]{KLY:16b}%
  \BibitemOpen
  \href@noop {} {} \bibinfo {note} {{The cross-correlations
  of ${\bm{\varphi }}$ and ${\bm{\theta}}$ are antisymmetric, see Appendix A,
  and thus do not contribute to RG flows of the scattering terms of
  Eq.~\eqref{pert}.}}\BibitemShut {Stop}%
\bibitem [{\citenamefont {Arora}\ \emph {et~al.}(1997)\citenamefont {Arora},
  \citenamefont {Babai}, \citenamefont {Stern},\ and\ \citenamefont
  {Sweedyk}}]{SVP:1}%
  \BibitemOpen
  \bibfield  {author} {\bibinfo {author} {\bibfnamefont {S}~\bibnamefont
  {Arora}}, \bibinfo {author} {\bibfnamefont {L}~\bibnamefont {Babai}},
  \bibinfo {author} {\bibfnamefont {J}~\bibnamefont {Stern}}, \ and\ \bibinfo
  {author} {\bibfnamefont {Z}~\bibnamefont {Sweedyk}},\ }\bibfield  {title}
  {\enquote {\bibinfo {title} {The hardness of approximate optima in lattices,
  codes, and systems of linear equations},}\ }\href@noop {} {\bibfield
  {journal} {\bibinfo  {journal} {J Comput Syst Sci}\ }\textbf {\bibinfo
  {volume} {54}},\ \bibinfo {pages} {317} (\bibinfo {year} {1997})}\BibitemShut
  {NoStop}%
\bibitem [{\citenamefont {Micciancio}(2014)}]{SVP}%
  \BibitemOpen
  \bibfield  {author} {\bibinfo {author} {\bibfnamefont {D.}~\bibnamefont
  {Micciancio}},\ }\bibfield  {title} {\enquote {\bibinfo {title} {The shortest
  vector in a lattice is hard to approximate to within some constant},}\
  }\href@noop {} {\bibfield  {journal} {\bibinfo  {journal} {SIAM J. Comput.}\
  }\textbf {\bibinfo {volume} {30}},\ \bibinfo {pages} {2008} (\bibinfo {year}
  {2014})}\BibitemShut {NoStop}%
\bibitem [{KLY({\natexlab{d}})}]{KLY:16c}%
  \BibitemOpen
  \href@noop {} {} \bibinfo {note} {{Such an instability can
  be due to an unreasonable original choice of the channels, e.g., electrons
  with opposite spins, when a WS suppresses charge current but has no impact on
  a spin current, leading to a finite conductivity of each original channel.
  However, a non-trivial situation emerges where the original choice is fixed
  by the boundary conditions on the leads, or spatial positions of the
  interacting channels, or different nature of them like in the fermion-boson
  case, etc}}\BibitemShut {NoStop}%
\bibitem [{\citenamefont {Choi}\ \emph {et~al.}(2015)\citenamefont {Choi},
  \citenamefont {Sivan}, \citenamefont {Rosenblatt}, \citenamefont {Heiblum},
  \citenamefont {Umansky},\ and\ \citenamefont {Mahalu}}]{Heiblum:15}%
  \BibitemOpen
  \bibfield  {author} {\bibinfo {author} {\bibfnamefont {H.~K.}\ \bibnamefont
  {Choi}}, \bibinfo {author} {\bibfnamefont {I.}~\bibnamefont {Sivan}},
  \bibinfo {author} {\bibfnamefont {A.}~\bibnamefont {Rosenblatt}}, \bibinfo
  {author} {\bibfnamefont {M.}~\bibnamefont {Heiblum}}, \bibinfo {author}
  {\bibfnamefont {V.}~\bibnamefont {Umansky}}, \ and\ \bibinfo {author}
  {\bibfnamefont {D.}~\bibnamefont {Mahalu}},\ }\bibfield  {title} {\enquote
  {\bibinfo {title} {Robust electron pairing in the integer quantum hall effect
  regime},}\ }\href@noop {} {\bibfield  {journal} {\bibinfo  {journal} {Nat
  Commun}\ }\textbf {\bibinfo {volume} {6}},\ \bibinfo {pages} {7435} (\bibinfo
  {year} {2015})}\BibitemShut {NoStop}%
\bibitem [{\citenamefont {Giamarchi}(2004)}]{Giamarchi}%
  \BibitemOpen
  \bibfield  {author} {\bibinfo {author} {\bibfnamefont {T}~\bibnamefont
  {Giamarchi}},\ }\href@noop {} {\emph {\bibinfo {title} {Quantum Physics in
  One Dimension}}}\ (\bibinfo  {publisher} {Clarendon Press},\ \bibinfo
  {address} {London},\ \bibinfo {year} {2004})\BibitemShut {NoStop}%
\bibitem [{\citenamefont {Wentzel}(1951)}]{Wentzel}%
  \BibitemOpen
  \bibfield  {author} {\bibinfo {author} {\bibfnamefont {Gregor}\ \bibnamefont
  {Wentzel}},\ }\bibfield  {title} {\enquote {\bibinfo {title} {The interaction
  of lattice vibrations with electrons in a metal},}\ }\href {\doibase
  10.1103/PhysRev.83.168} {\bibfield  {journal} {\bibinfo  {journal} {Phys.
  Rev.}\ }\textbf {\bibinfo {volume} {83}},\ \bibinfo {pages} {168--169}
  (\bibinfo {year} {1951})}\BibitemShut {NoStop}%
\bibitem [{\citenamefont {Bardeen}(1951)}]{Bardeen:51}%
  \BibitemOpen
  \bibfield  {author} {\bibinfo {author} {\bibfnamefont {J}~\bibnamefont
  {Bardeen}},\ }\bibfield  {title} {\enquote {\bibinfo {title}
  {Electron-vibration interactions and superconductivity},}\ }\href@noop {}
  {\bibfield  {journal} {\bibinfo  {journal} {Rev. Mod. Phys.}\ }\textbf
  {\bibinfo {volume} {23}},\ \bibinfo {pages} {261} (\bibinfo {year}
  {1951})}\BibitemShut {NoStop}%
\end{thebibliography}


\end{document}